\documentclass[%        	Class options:
aps,%                   	American Physical Society
prc,%                   	Physical Review D
twocolumn,%                     twocolumn
showpacs,%              	Displays PACS after abstract
superscriptaddress,%    	Authors' addresses linked with superscripts
nofootinbib,%           	Does not treat footnotes as references
floatfix]%              	Fixes float errors
{revtex4}%              	REVTEX 4 Package used
\usepackage{graphicx,%  	Default Latex 2eps package for embedding figures
longtable}%             	Useful for long table

\usepackage{bm}

\def\be{\begin{equation}}
\def\ee{\end{equation}}
\def\dg{\dagger}
\def\bee{\begin{eqnarray}}
\def\eee{\end{eqnarray}}

\newcommand{\lf}{\left(}
\newcommand{\rh}{\right)}

 \newcommand{\gd}[1]{\gamma_{#1}}

\begin{document}

%======================================================================================
\title{The light neutrino exchange mechanism of the $0\nu\beta\beta$-decay with left- and
  right-handed leptonic and hadronic currents revisited.}
%--------------------------------------------------------------------------------------
%
%
\author{Du\v{s}an \v Stef\'{a}nik}
\affiliation{Department of Nuclear Physics and Biophysics, Comenius
University, Mlynsk\'{a} dolina F1, SK-842 48
Bratislava, Slovakia}
\author{Rastislav Dvornick\'{y}}
\affiliation{Department of Nuclear Physics and Biophysics, Comenius
University, Mlynsk\'{a} dolina F1, SK-842 48
Bratislava, Slovakia}
\affiliation{Dzhelepov Laboratory of Nuclear Problems, JINR 141980 Dubna,
Russia}
\author{Fedor \v{S}imkovic}
\affiliation{Department of Nuclear Physics and Biophysics, Comenius
University, Mlynsk\'{a} dolina F1, SK-842 48
Bratislava, Slovakia}
\affiliation{Boboliubov Laboratory of Theoretical Physics, JINR 141980 Dubna,
Russia}
\affiliation{Czech Technical University in Prague, 128-00 Prague, Czech Republic}
\author{Petr Vogel}
\affiliation{Kellogg Radiation Laboratory, California Institute of Technology,
Pasadena, CA 91125, USA} 

\begin{abstract}%.......................................................................
The extension of Majorana neutrino mass mechanism of the neutrinoless double-beta decay
($0\nu\beta\beta$) with the inclusion of right-handed leptonic and hadronic currents
is revisited. While only the exchange of light neutrinos is assumed, the $s_{1/2}$
and $p_{1/2}$-states of emitted electrons as well as recoil corrections to the nucleon
currents  are taken into account. Within the standard approximations the decay rate
is factorized into a sum of products of kinematical phase-space factors, nuclear matrix
elements and the fundamental parameters that characterize the lepton number violation.
Unlike in the previous treatments the induced pseudoscalar term of hadron current
is included, resulting in additional nuclear matrix elements.
%The phase-space factors are expressed using the Dirac components of the electron
%radial wave functions in the $s_{1/2}$ and $p_{1/2}$ wave states. 
An improved numerical computation  of the phase-space factors is presented, based on
the exact Dirac wave functions of the $s_{1/2}$ and  $p_{1/2}$ electrons with finite nuclear
size and electron screening taken into account. The dependence of values
of these phase-space factors on the different approximation schemes used in evaluation
of electron wave functions is discussed.
The upper limits for effective neutrino mass and the parameters $\langle\lambda\rangle$
and $\langle\eta\rangle$ characterizing the right-handed current mechanism
are deduced from data on the $0\nu\beta\beta$-decay of $^{76}$Ge and $^{136}$Xe
using nuclear matrix elements calculated within the nuclear shell model
and quasiparticle random phase approximation.
The differential decay rates, i.e. the angular correlations and the single electron energy
distributions for various combinations of the total lepton number violating parameters
that can help to disentangle the possible mechanism are described and discussed. 
\end{abstract}%.........................................................................
\medskip

\maketitle

%%%%%%%%%%%%%%%%%%%%%%%%%%%%%%%%%%%%%%%%%%%%%%%%%%%%%%%%%%%%%%%%%%%%%%%%%%%%%%%%%%%%%%%%%%%%%%%%%%%%
%
\section{Introduction}
%
%%%%%%%%%%%%%%%%%%%%%%%%%%%%%%%%%%%%%%%%%%%%%%%%%%%%%%%%%%%%%%%%%%%%%%%%%%%%%%%%%%%%%%%%%%%%%%%%%%%%

The neutrinoless double-beta ($0\nu\beta \beta$) decay is a process in which an atomic nucleus with Z protons decays to another
one with two more protons and the same mass number A, by emitting two
electrons and nothing else
\bee
(A,Z)\rightarrow(A,Z+2)+2e^-.
\eee

Observing the $0\nu \beta \beta$-decay guaranties that neutrinos are massive Majorana
particles - it means that neutrino is identical to its own antiparticle \cite{schechter}.
$0\nu \beta \beta$-decay violates total lepton
number conservation and is forbidden in the Standard Model.

When the light-neutrino exchange produced by left-handed currents is
the driving mechanism for $0\nu\beta\beta$-decay, the relation between the effective
Majorana neutrino mass and the inverse half-life  of the $0\nu\beta\beta$-decay 
can be written as \cite{review}
\be \label{facbez}
\left[T_{1/2}^{0\nu} \right]^{-1}=G^{0\nu}(Q,Z)g_A^4\left|M^{0\nu} \right|^2 \frac{\left|m_{\beta\beta}\right|^2}{m_e^2},
\ee
where $G^{0\nu}(Q,Z)$, $g_A$ and $M^{0\nu}$ represent the phase-space factor,
the axial-vector coupling constant and the nuclear matrix element  of the process, respectively.
In that case the ultimate goal of the search for
$0\nu\beta\beta$-decay is the determination of the effective
Majorana neutrino mass,
\begin{equation}
m_{\beta\beta} = U^2_{e 1} m_1 + U^2_{e 2} m_2 + U^2_{e 3} m_3.
\label{mbb}
\end{equation}
Here, $U_{e i}$ and $m_i$ (i=1,2,3) are elements of
Pontecorvo-Maki-Nakagawa-Sakata neutrino mixing matrix and masses
of neutrinos, respectively.

An improved calculation of $G^{0\nu}$ in that case, taking into account the electron
Dirac wave functions with finite nuclear size and electron screening, was performed
in \cite{iachello}. The main theoretical uncertainty
is represented in the computed values of the nuclear matrix elements. There is a factor 
of 2-3 difference between the different methods of calculations of the $M^{0\nu}$.

The left-right symmetric theories \cite{Pati,Mohap} provide a natural framework to understand the origin of neutrino 
Majorana masses. In general one cannot predict the scale where the left-right symmetry is realized, but 
it is natural to assume that it is as low as $\sim$ a few TeV which can affect the $0\nu\beta\beta$ decay rate
significantly.

In the left-right symmetric theories in addition to the left-handed
V-A weak currents also leptonic and hadronic right-handed V+A weak currents are present.
In that case new mechanism of the $0\nu\beta\beta$-decay need to be considered. 
In the past the  $0\nu\beta\beta$-decay rate in the presence of the
right-handed leptonic and hadronic currents was discussed in \cite{Doies,tomoda}. 
Recently, contributions to the $0\nu\beta\beta$-decay in a TeV-scale left-right symmetric
models for type-I seesaw dominance were revisited \cite{Tello,Nemevsek,Barry13,Dev}.
By making a  qualitative analysis without considering relevant phase-space factors and nuclear matrix elements it was
found that $W_L$-$W_R$ exchange ($\lambda$ mechanism) and $W_L$-$W_R$ mixing ($\eta$ mechanism)
could give dominant contribution to the $0\nu\beta\beta$-decay amplitude by assuming
a wide particle physics available parameter space including left-right neutrino mixing
\cite{Barry13,Dev}. We note that the discovery of a non-standard $0\nu\beta\beta$-decay mechanism
such as a right-handed current would rule out most models of baryogenesis at scales above 40 TeV \cite{deppisch}.

The goal of this paper is to revisit the $0\nu\beta\beta$-decay mechanism due to
the right-handed currents by considering exact Dirac wave function of electrons and the
higher order terms of nucleon current. We note that in \cite{Doies} the effect of
induced pseudoscalar term of nucleon current was neglected and phase space factors
were expressed using approximate electron wave functions for a uniform 
charge distribution in a nucleus by keeping only the lowest terms in the power expansion
in r. In that context the subject of interest is the comparison of the power
expansion versus the exact treatment and the finite nuclear size effects. 
In this work the newly derived decay rate will be written in a compact form  and the corresponding
nuclear matrix elements will be presented by assuming the usual closure approximation for
intermediate nuclear states. The phase-space factors will be evaluated by
using the exact Dirac wave functions with finite nuclear size  and electron screening.
The differential characteristics,  i.e. the angular correlations and the single electron energy distributions will
be described and discussed, and the  decay rates and updated constraints
on the lepton number violating parameters for different combinations of the total lepton
number violating parameters will be recalculated. This will make it possible to judge
the importance of the $\lambda$ and $\eta$ mechanism.

%%%%%%%%%%%%%%%%%%%%%%%%%%%%%%%%%%%%%%%%%%%%%%%%%%%%%%%%%%%%%%%%%%%%%%%%%%%%%%%%%%%%%%%%%%%%%%%%%%%%
%
\section{Electron wave functions}
\label{Sec:ewf}
%
%%%%%%%%%%%%%%%%%%%%%%%%%%%%%%%%%%%%%%%%%%%%%%%%%%%%%%%%%%%%%%%%%%%%%%%%%%%%%%%%%%%%%%%%%%%%%%%%%%%%

An important ingredient in the calculation of the electron energy spectrum
is the radial electron wave function distorted by the Coulomb field.
We adopt the Dirac wave functions in a central field,
\begin{eqnarray}
\Psi (\varepsilon,\bm{r}) = \sum_{\kappa \mu}
   \left( \begin{array}{c}
             g_{\kappa}(\varepsilon,r) \chi_{\kappa \mu} (\hat{r}) \\
	   i f_{\kappa}(\varepsilon,r) \chi_{-\kappa \mu} (\hat{r})
          \end{array}
   \right),    
\end{eqnarray}
given, e.g. by Rose \cite{rose}. Here, $\varepsilon$ and $\bm{r}$ stand for
energy and position vector of the electron, respectively,  $r = |\bm{r}|$ and
$\hat{r} = \bm{r}/r$. The index $\kappa$ takes positive and 
negative integer values ($\kappa=\pm k;~k=1,2,3 \cdots$). Total angular momentum is
given as $j_{\kappa}=|\kappa|-1/2$ while orbital angular momentum takes values
\begin{eqnarray}
l_\kappa &=& |\kappa|-1 \hspace{0.6cm} \rm{for} \hspace{0.6cm} \kappa < 0\nonumber\\
        &=& \kappa      \hspace{1.4cm} \rm{for} \hspace{0.6cm} \kappa > 0
\end{eqnarray}
Radial wave functions $g_{\kappa}(\varepsilon,r)$ and $f_{\kappa}(\varepsilon,r)$
obey the radial Dirac equations,
\begin{eqnarray}
\frac{d g_{\kappa}}{dr} &=& -\frac{\kappa +1}{r} g_{\kappa} 
                        + (\varepsilon-V(r)+m_e) f_{\kappa} \nonumber\\			 
\frac{d f_{\kappa}}{dr} &=& -(\varepsilon-V(r)-m_e) g_{\kappa}
                        +\frac{\kappa -1}{r} f_{\kappa},                   
\end{eqnarray}
where $V(r)$ is the central Coulomb potential. The natural
units $\hbar=c=1$ are used.

The electron wave function expressed in terms of spherical waves
is given by
\begin{eqnarray}
\Psi(\varepsilon,\bm{r}) &=& \Psi^{(s_{1/2})}(\varepsilon,\bm{r})  
 + \Psi^{(p_{1/2})}(\varepsilon,\bm{r})  + \cdots .
\end{eqnarray}
Here, superscript displays the orbital angular momentum 
($l_{\kappa}=0,1,2, \cdots$) in a spectroscopic notation
($l_{\kappa}=s,p,d,  \cdots$) and the total angular momentum $j_{\kappa}$.
The states of particular interest in our calculations are:
\begin{eqnarray}
\Psi^{(s_{1/2})}(\varepsilon,\bm{r}) &=& 
\left(\begin{array}{l}
g_{-1 }(\varepsilon,r)\chi_s \\
f_{+1 }(\varepsilon,r) (\mathbf{\sigma\cdot\hat{p}}) \chi_s
\end{array}\right), \nonumber\\
\Psi^{(p_{1/2})} (\varepsilon,\bm{r}) 
&=&  i
\left( \begin{array}{l}
 g_{+1}(\varepsilon, r)~ (\mathbf{\sigma}\cdot\hat{\mathbf{r}})~ 
(\mathbf{\sigma}\cdot\hat{\mathbf{p}})~ \chi_s\\
- f_{-1}(\varepsilon, r)~ (\mathbf{\sigma}\cdot\hat{\mathbf{r}})~ \chi_s
\end{array}
\right),\nonumber\\
\end{eqnarray}
where $\hat{p}=\bm{p}/p$ and $p$ is the electron momentum.
The asymptotic behavior of the radial wave functions for large 
values of $p r$ is given by
\begin{eqnarray}
 &&\left(
   \begin{array}{c}
      g_{\kappa}(\varepsilon,r) \\
      f_{\kappa}(\varepsilon,r)
   \end{array}
 \right)\\
&&\approx \frac{1}{p r}
 \left( 
   \begin{array}{c}
      \sqrt{\frac{\varepsilon+m_e}{2\varepsilon}} \sin (p r-l\frac{\pi}{2}
      +\delta_k+\alpha Z_f \frac{\varepsilon}{p} \log 2p r) \\
      \sqrt{\frac{\varepsilon-m_e}{2\varepsilon}} \cos (p r-l\frac{\pi}{2}
      +\delta_k+\alpha Z_f \frac{\varepsilon}{p} \log 2p r) 
   \end{array}  
 \right).\nonumber 
\end{eqnarray}
$Z_f$ is the charge of the final system which generates the
potential $V(r)$.

In what follows, different approximation schemes for the
calculation of radial wave functions $g_{\pm 1}$ and $f_{\pm 1}$
associated with emitted electron in the $s_{1/2}$ and $p_{1/2}$
wave states are briefly presented.\\

{\it The  approximation scheme A):} The relativistic electron wave function in a uniform 
charge distribution in nucleus is considered. The lowest terms in the power expansion
in r are taken into account. The radial wave functions take the form 
\begin{eqnarray}
 \left(
   \begin{array}{c}
      g_{-1}(\varepsilon,r) \\
      f_{+1}(\varepsilon,r)
   \end{array}
 \right)
&\approx& 
\sqrt{F_0(Z_f,\varepsilon)} 
\left(
   \begin{array}{c}
      \sqrt{\frac{\varepsilon+m_e}{2\varepsilon}} \\
      \sqrt{\frac{\varepsilon-m_e}{2\varepsilon}}
   \end{array}
 \right), \nonumber\\
 \left(
   \begin{array}{c}
      g_{+1}(\varepsilon,r) \\
      f_{-1}(\varepsilon,r)
   \end{array}
 \right)
&\approx& 
\sqrt{F_0(Z_f,\varepsilon)}\times\nonumber\\ 
&&\left(
   \begin{array}{c}
      \sqrt{\frac{\varepsilon-m_e}{2\varepsilon}} [\alpha Z_f/2 +(\varepsilon+m_e)r/3]\\
     -\sqrt{\frac{\varepsilon+m_e}{2\varepsilon}} [\alpha Z_f/2 +(\varepsilon-m_e)r/3]
   \end{array}
 \right). \nonumber\\
\end{eqnarray}
Here, $F_{k-1}$ (for $k=1,2, \cdots$) is given by
\begin{eqnarray}
F_{k-1} &=& \left[ \frac{\Gamma(2k+1)}{\Gamma(k)\Gamma(1+2\gamma_{k})}
\right]^2 (2 p r)^{2(\gamma_{k}-k)} e^{\pi y} \nonumber\\
&& \times \Gamma(\gamma_{k}+iy)  \mid^2,
\end{eqnarray}
where 
\begin{eqnarray}
\gamma_{k} &=& \sqrt{k^2-(\alpha Z_f)^2} \nonumber\\
y &=& \alpha Z_f \frac{\varepsilon}{p}.
\end{eqnarray}
This approximation scheme was commonly used in the past calculations of the phase-space integrals
of double beta decay processes \cite{Doies}. 

{\it The  approximation scheme B):} The analytical solution of the Dirac equation for
the point-like nucleus is considered \cite{landau}. Radial wave functions take then the form

\begin{eqnarray}
g_{\kappa}(\varepsilon,r) &=& \frac{\kappa}{k}
 \frac{1}{p r} \sqrt{\frac{\varepsilon+m_e}{2\varepsilon}}
       \frac{|\Gamma(1+\gamma_k+iy)|}{\Gamma(1+2\gamma_k)} 
       (2p r)^{\gamma_k} e^{\pi y/2}\nonumber\\
       && \Im \left\{ e^{i(p r+\xi)}~
        _1F_1(\gamma_k-iy,1+2\gamma_k,-2ip r) \right\}, \nonumber\\
f_{\kappa}(\varepsilon,r) &=& \frac{\kappa}{k}
\frac{1}{p r} \sqrt{\frac{\varepsilon-m_e}{2\varepsilon}}
       \frac{|\Gamma(1+\gamma_k+iy)|}{\Gamma(1+2\gamma_k)} 
       (2p r)^{\gamma_k} e^{\pi y/2}\nonumber\\
       && \Re \left\{ e^{i(p r+\xi)}~
        _1F_1(\gamma_k-iy,1+2\gamma_k,-2ip r) \right\}, \nonumber\\     
\end{eqnarray}     

with 
\begin{eqnarray}
e^{-2i\xi} = \frac{\gamma_k-iy}{\kappa-iym_e/\varepsilon}.
\end{eqnarray}
Here,  $_1F_1(a,b,z)$ is the confluent hypergeometric function. 

{\it The  approximation scheme C):} The exact Dirac wave functions
with finite nuclear size, which is taken into account
by a uniform charge distribution in a sphere of nuclear
radius $R$, are assumed \cite{iachello}. The numerical calculation can be
accomplished by the subroutine package RADIAL \cite{radial}, where 
the input central potential is given by
\begin{eqnarray}
 V(r) = \left\{ 
           \begin{array}{lr}
	    -\frac{\alpha Z_f}{2R} \left(3-\left( \frac{r}{R} \right)^2 \right)  
	   & \rm{for} \hspace{0.6cm} r\leq R \\
	   -\frac{\alpha Z_f}{r} & \hspace{1cm} \rm{for} \hspace{0.6cm} r > R 
	   \end{array}
	  \right.
	  \label{eq-V0}
\end{eqnarray}
Here, $R=r_0A^{1/3}$ with $r_0=1.2$ fm.
In the code the radial Dirac equations are solved by using piecewise
exact power series expansion, which are summed up to a prescribed accuracy 
so that truncation errors can be avoided completely.

\begin{center}
  \begin{figure}[!t]
  \includegraphics[width=1.10\columnwidth]{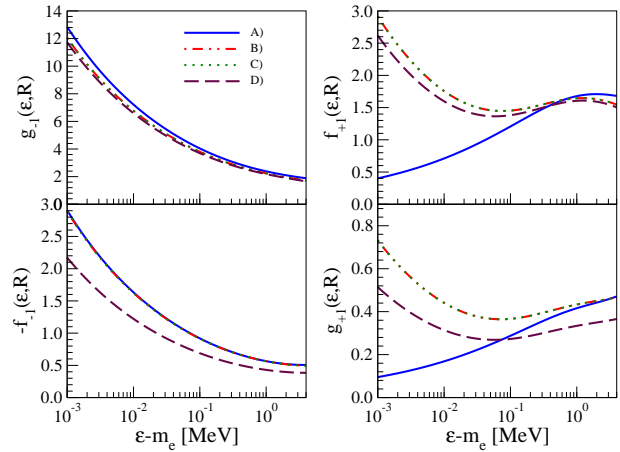}
  \caption{The radial wave functions of an electron in the $s_{1/2}$ wave state,
    $g_{-1}(\varepsilon )$ and $f_{+1}(\varepsilon )$ (upper panels), and
    in the $p_{1/2}$ wave state,  $g_{+1}(\varepsilon )$, $f_{-1}(\varepsilon )$ (lower panels),
    as function of the electron kinetic energy $\varepsilon - m_e$. Results are presented for an electron emitted 
    in the double $\beta$-decay of $^{150}$Nd. In an evaluation of radial wave functions (w.f.)
    four different approximation schemes are considered (see Sec.(\ref{Sec:ewf}) for details):
    A) The standard approximation of Doi et al. \cite{Doies};
    B) An analytical solution of Dirac equations for a pointlike nucleus is assumed;
    C) An  exact solution of Dirac equations for a uniform charge distribution in
    nucleus is considered at nuclear surface; D) The same as the previous case but the electron
    screening is taken into account \cite{iachello}.} 
\label{fig:EWF}
\end{figure}
\end{center}

{\it The  approximation scheme D):} The exact Dirac wave functions
with finite nuclear size and electron screening are used \cite{iachello}. 
The effect  of screening of atomic electrons is taken into account
by the Thomas-Fermi approximation. It uses the solution of the Thomas-Fermi equation,
\begin{eqnarray}
\frac{d^2 \varphi}{d x^2} = \frac{\varphi^{3/2}}{\sqrt{x}},
\label{eq-TF}
\end{eqnarray}
with $x=r/b$, where
\begin{eqnarray}
b = \frac{1}{2} \left( \frac{3\pi}{4} \right)^{2/3}
     a_0 Z_f^{-1/3}.
\end{eqnarray}
The Thomas-Fermi function can be rewritten in terms of
an effective charge $\varphi(x)=Z_{eff}(x)/Z_f$. Therefore,
boundary conditions 
\begin{eqnarray}
\varphi(0)=1, \hspace{1cm} \varphi(\infty)=\frac{2}{Z_f}
\end{eqnarray}
of  Eq.(\ref{eq-TF}) take into account the fact 
that the final atom is a positive ion with electric charge $+2$. 
We adopt here the Majorana method described in Ref. \cite{esposito} 
in order to solve Eq.(\ref{eq-TF}). The input potential is then 
$V(r)=\varphi(r)V_0(r)$, where $V_0(r)$ is defined in Eq. (\ref{eq-V0}).

In Fig. \ref{fig:EWF} radial wave functions of an electron in the $s_{1/2}$ wave state
($g_{-1}(\varepsilon )$ and $f_{+1}(\varepsilon )$) and in the $p_{1/2}$ wave state
($g_{+1}(\varepsilon )$, $f_{-1}(\varepsilon )$)
evaluated at $r = R$ are plotted as a function of the electron
kinetic energy $\varepsilon - m_e$ for the case of  the double $\beta$-decay of $^{150}$Nd.
We see that wave functions A, which correspond to leading finite-size Coulomb,
agree qualitatively well with wave functions of the other approaches in the case
of $g_{-1}(\varepsilon )$ and $f_{-1}(\varepsilon )$ but that there are significant differences
for $g_{+1}(\varepsilon )$ and $f_{+1}(\varepsilon )$. These differences are apparent
especially at lower energies. We notice also a rather good agreement between results
for wave functions of B and C in general.  The screening of atomic electrons
affects mostly the $p_{1/2}$ wave functions, but is essentially negligible for the $s_{1/2}$ states.

%%%%%%%%%%%%%%%%%%%%%%%%%%%%%%%%%%%%%%%%%%%%%%%%%%%%%%%%%%%%%%%%%%%%%%%%%%%%%%%
%
\section{Decay rate for the neutrinoless double-beta decay}
%
%%%%%%%%%%%%%%%%%%%%%%%%%%%%%%%%%%%%%%%%%%%%%%%%%%%%%%%%%%%%%%%%%%%%%%%%%%%%%%%

One of the most prominent new physics model that incorporates the LNV
  and which leads to potentially observable rates for the $0\nu\beta\beta$-decay
  is the minimal left-right symmetric model (LRSM) \cite{Pati,Mohap} which extends the
  standard model gauge symmetry to the group $SU(2)_L\otimes SU(2)_R \otimes U(1)_{B-L}$.
  The right-handed neutrinos necessary appear here as a part of the $SU(2)_R$ doublets.
  The lepton multiplets $L_i=(\nu_i,l_i)$ are characterized by the quantum numbers
  $Q_{L_L}=(1/2,0,-1)$ and $Q_{L_R}=(0,1/2,-1)$ under $SU(2)_L\otimes SU(2)_R \otimes U(1)_{B-L}$.
  The Higgs sector contains a bidoublet $\phi$ and two triplets $\Delta_L$ and $\Delta_R$
  with vacuum expectation values (VEV) $v_L$ and $v_R$, respectively. The VEVs fulfill
  the condition $v_Lv_R=v^2$. The VEV  $v_R$  breaks $SU(2)_R \otimes U(1)_{B-L}$ to $U(1)_{Y}$
  and generates masses for the right-handed $W_R$ and $Z_R$ gauge bosons, and the heavy neutrinos.
  The $W_L$ and $W_R$ are in general not mass eigenstates and are related to the mass eigenstates
  $W_1$ and $W_2$ with masses $M_1$ and $M_2$ ($M_1<M_2$) as 
\bee
 \left(
\begin{array}{l}
W_L^-  \\
 W_R^-  \\
 \end{array}
\right)&=& \left(
\begin{array}{ll}
\cos{\zeta}&\sin{\zeta} \\
 -\sin{\zeta}&\cos{\zeta} \\
 \end{array}
\right)\left(\begin{array}{l}
W_1^-  \\
 W_2^-  \\
 \end{array}
\right).
 \nonumber\\
\label{maticaW}
 \eee
Then, the effective current-current interaction which can trigger the $0\nu\beta\beta$-decay 
can be written as \cite{Doies}
\bee \label{hamilweakrh}
H^\beta &=& \frac{G_{\beta}}{\sqrt{2}} ~\left[
j_L^{~\rho}J^{\dg}_{L\rho }+\chi j_L^{~\rho}J^{\dg}_{R\rho } \right.\nonumber\\
&+&\left.
~\eta j_R^{~\rho}J^{\dg}_{L\rho } +\lambda j_R^{~\rho}J^{\dg}_{R\rho } +  h.c.\right].
\eee
Here, $G_{\beta}=G_F\cos{\theta_C}$, where $G_F$ and $\theta_C$ are Fermi constant and Cabbibo angle, respectively.
The coupling constants $\lambda$, $\eta$ and $\chi$ are chosen to be real. We have
\bee \label{relacieLRSM}
\eta &\simeq& -\tan{\zeta},\quad \chi=\eta, \nonumber \\
\lambda &\simeq&  (M_{W_1}/M_{W_2})^2.
\eee
The left- and right-handed leptonic currents are given by 
\bee
j_L^{~\rho}=\bar{e}\gd{\rho}(1-\gd{5})\nu_{eL}, \qquad j_R^{~\rho}=\bar{e}\gd{\rho}(1+\gd{5})\nu_{eR}.\nonumber \\
\eee
The $\nu_{eL}$ and  $\nu_{eR}$ are the weak 
eigenstate electron neutrinos, which are expressed as superpositions of the light and heavy 
 mass eigenstate Majorana neutrinos   
$\nu_j$ and $N_j$, respectively. The electron neutrinos eigenstates can be expressed as
\bee\label{lambdaeta}
&&\nu_{eL}=\sum_{j=1}^3\lf U_{ej}\nu_{jL}+S_{ej} (N_{jR})^C \rh, \nonumber \\
&& \nu_{eR}=\sum_{j=1}^3 \lf T_{ej}^*(\nu_{jL})^C+V_{ej}^*N_{jR} \rh.
\eee

The (3 + 3) scenario is assumed. The $3\times 3$ block matrices in  flavor  space $U, S, T, V$, generalizations of the
Pontecorvo-Maki-Nakagawa-Sakata matrix, constitute the $6\times6$ unitary neutrino mixing matrix \cite{Xing}
\bee
\mathcal{U}&=& \left(
\begin{array}{ll}
U & S\\
 T &V \\
 \end{array}
\right), 
\label{maticazmies}
\eee
which diagonalizes the general $6\times6$ neutrino mass matrix in the basis $(\nu_L,(N_R)^C)^T$:
\bee
 \mathcal{M} &=& \left(
\begin{array}{ll}
M_L & M_D \\
 M_D^T & M_R \\
 \end{array}
\right)
\label{maticanut}
 \eee
 with Majorana and Dirac mass terms, which are proportional to $M_L\approx y_M v_L,~M_R\approx y_M v_R$ and $M_D\approx y_D v$, 
 where $y_M$ and $y_D$ are the Yukawa couplings. The full parametrization of matrix $\mathcal{U}$ includes 15 rotational
 angles and 10 Dirac and 5 Majorana CP violating phases. It is possible to decompose U as follows \cite{Xing}
\bee
 {\mathcal{U}} &=& \left(
\begin{array}{ll}
\mathbf{1} & \mathbf{0}\\
 \mathbf{0} & U_0\\
 \end{array}
\right)
\left(
\begin{array}{ll}
A & R\\
S & B\\
 \end{array}
\right)
\left(
\begin{array}{ll}
V_0 & \mathbf{0}\\
 \mathbf{0} & \mathbf{1}\\
 \end{array}
\right),
\label{maticaXing}
 \eee
 where $\mathbf{0}$ and $\mathbf{1}$ are the $3\times 3$ zero and identity matrices, respectively. 
 The parametrization of matrices
 A, B, R and S and corresponding orthogonality relations are given in \cite{Xing}. In the limit case
 A = $\mathbf{1}$, B =$\mathbf{1}$, R = $\mathbf{0}$ and S = $\mathbf{0}$ there would be a separate mixing of
 heavy and light neutrinos, which would participate only in left and right-handed currents, respectively. In that
 case only the neutrino mass mechanism of the $0\nu\beta\beta$-decay would be allowed and exchange
 neutrino momentum dependent mechanisms associated with the $W_L$-$W_R$ exchange and $W_L$-$W_R$ mixing would be forbidden.
 If masses of heavy neutrinos are above the TeV scale, the mixing angles responsible for mixing of light and
 heavy neutrinos are small. By neglecting the mixing between different generations of light and heavy neutrinos
 A, B, R and S matrices can be approximated as follows:
\begin{eqnarray}
  A \approx \mathbf{1}, B \approx \mathbf{1}, R \approx \frac{m_D}{m_{LNV}}\mathbf{1}, S \approx -\frac{m_D}{m_{LNV}}\mathbf{1}.
\label{whynot}
\end{eqnarray}
Here, $m_D$ represents energy scale of charge leptons and $m_{LNV}$ is the total lepton number violating scale,
which corresponds to masses of heavy neutrinos.
For sake of simplicity the same mixing angle is assumed for each generation of mixing of light and heavy neutrinos.
We see that $U_0$ can be identified to a good approximation with the PMNS matrix and $V_0$ is its analogue for
heavy neutrino sector. Since $V_0$ is unknown, it is common to assume that the structure of
$V_0$ is the same one as $U_0$. 

Assuming the non-relativistic impulse approximation
the left and right hadronic currents  $J^{\rho\dg}_{L}$ and $J^{\rho\dg}_{R}$ become
\cite{Doies}
\bee \label{nonhad}
&&J^{\rho}_{L}(\mathbf{x})=\sum_{n}\tau_{n}^+\delta(\mathbf{x}-\mathbf{r}_n)
 \left[\left(g_{V}-g_{A}C_n\right)g^{\rho 0} \right.\nonumber\\
&&+\left.g^{\rho k}\left(g_{A}\sigma_n^k-g_{V}D_n^k-g_{P}~q_n^{k}~
\frac{\vec{\sigma}_n \cdot \mathbf{q}_n}{2m_N}
\right)\right]~, \nonumber \\
&&J^{\rho}_{R}(\mathbf{x})=\sum_{n}\tau_{n}^+\delta(\mathbf{x}-\mathbf{r}_n)
 \left[ \left(g_{V}^{\prime}+g_{A}^{\prime}C_n\right)g^{\rho 0} \right.\nonumber\\
&&+\left.g^{\rho k}\left(-g_{A}^{\prime}\sigma_n^k-g_{V}^{\prime}D_n^k+g_{P}^{\prime}~q_n^{k}~
\frac{\vec{\sigma}_n \cdot \mathbf{q}_n}{2m_N}
\right)\right]~. \nonumber \\
\eee
Here, $\mathbf{q}_n=\mathbf{p}_n-\mathbf{p}_n^{'}$ is the momentum transfer between the nucleons.
The final proton (initial neutron) possesses  energy $E_n'$ ($E_n$)
and momentum $\mathbf{p}_n'$ ($\mathbf{p}_n$). $\vec{\sigma}_n$, $\tau^+_n$
and $\mathbf{r}_n$ are the Pauli matrix, the isospin raising operator 
and the position operator, respectively. These operators act on the $n$-th nucleon.

The nucleon recoil operators $C_n$ and $\mathbf{D}_n$ are given by
\bee
&&C_n=\frac{\vec{\sigma} \cdot \left(\mathbf{p}_n+\mathbf{p}_n^{'}\right)}{2m_N}-\frac{g_P}{g_A}\left(E_n-E_n^{'}\right)
\frac{\vec{\sigma} \cdot \mathbf{q}_n}{2m_N},\nonumber\\
&&\mathbf{D}_n=\frac{\left(\mathbf{p}_n+\mathbf{p}_n^{'}\right)}{2m_N}-i\left(1+\frac{g_M}{g_V}\right)
\frac{\vec{\sigma} \times \mathbf{q}_n}{2m_N}.
\eee
Here,  $m_N$ is the nucleon mass. $q_V\equiv q_V(q^2)$, $q_M\equiv q_M(q^2)$, $q_A\equiv q_A(q^2)$ and $q_P\equiv q_P(q^2)$
are, respectively, the vector, weak-magnetism, axial-vector and induced pseudoscalar form-factors in the case
of left-handed hadronic currents. As the strong and electromagnetic interactions conserves parity 
there are relations among form-factors entering the left-handed and right-handed hadronic currents
\cite{Doies}:
\be
\frac{g_A}{g_V}=\frac{g_A^{\prime}}{g_V^{\prime}}, \qquad \frac{g_M}{g_V}=\frac{g_M^{\prime}}{g_V^{\prime}}, \qquad
 \frac{g_P}{g_V}=\frac{g_P^{\prime}}{g_V^{\prime}}.
 \ee
We note that the induced pseudoscalar term of the space component of  hadronic currents was not
taken into account in derivation of the $0\nu\beta\beta$-decay rate presented in \cite{Doies}.
This simplification is avoided here. 

Due to helicity matching of the propagating neutrino 
the decay amplitude can be divided into two parts:\\
a) If both vertices are $V-A$ or $V+A$, the amplitude of the process is proportional to the neutrino mass
$m_j$. We shall denote the corresponding parts of the $0\nu\beta\beta$-decay amplitude $L$-$L$ and $R$-$R$
terms, respectively.\\
b) If one vertex is $V-A$ and the other is $V+A$, the four momentum of propagating neutrino
$q^{\mu}=(\omega, \mathbf{q})$ contributes. The corresponding part of the amplitude, which is
denoted as $L$-$R$, is further separated into two terms, the $\omega$-term and the $\mathbf{q}$-term.\\
In the case of $L$-$L$- and $R$-$R$-terms the dominant contribution  is associated with the emission
of electrons in the $s_{1/2}$-wave state \cite{Stefanik}. However, the $\mathbf{q}$-term
changes the parity and therefore it requires that one of the final electrons to be in $s_{1/2}$-wave
while the other to be in $p_{1/2}$-wave, or both electrons are in the $s_{1/2}$-wave and nucleon recoil operator
is taken into account. Nevertheless, the $\mathbf{q}$-term is not negligible since
the $\omega$-term is suppressed by a factor $\varepsilon_{12}/q\approx 1/40$ \cite{Doies},
that makes the $\mathbf{q}$-term comparable or even larger in comparison with  the $\omega$-term.

The standard approximations of Ref. \cite{Doies} are adopted:\\
i)  Only mechanisms with the exchange of light neutrinos are considered and contributions from heavier neutrinos
 are neglected. Recently, it was concluded in Ref. \cite{Barry13,Dev} that mechanisms with the exchange of light neutrinos could give dominant
contributions to the $0\nu\beta\beta $ amplitude in a wide range of the LRSM parameter space.\\
ii) Closure approximation.
Within this approximation energies of intermediate nuclear states 
$E_n -(E_i+E_f)/2$ are replaced by an average of 
$\bar{E}_n-(E_i+E_f)/2 \sim 10$ MeV and the sum over intermediate states
is taken by closure, $\sum_n |n\rangle\langle n| = 1$.\\
iii) The $R$-$R$-part of the amplitude, that is 
multiplied by factor $|\lambda^2\sum_{j}m_jT_{ej}^{*2}|$, becomes negligible in comparison with $m_{\beta\beta}$.
Thus it is neglected. \\ 
iv) The terms proportional to the square of the nucleon recoil operators are also neglected.\\
v) For $L$-$L$-part of amplitude only electrons  in the  $s_{1/2}$ wave state are included. 
The inclusion of the $p_{1/2}$ electrons leads only to negligible contribution to
the $0\nu\beta\beta$ standard decay rate \cite{Stefanik}.\\
vi) In the case of the  $L$-$R$ term, two-nucleon potentials associated with the spatial $q$ and time $\omega$
components of neutrino exchange potentials are simplified as follows:
\bee
H_{q}^l(\mathbf{x})&=&\int\frac{d\mathbf{q}}{2\pi^2}\lf\frac{q^l}{q+\Delta-\varepsilon_{12}}+\frac{q^l}
{q+\Delta+\varepsilon_{12}} \rh e^{i\mathbf{q}\cdot \mathbf{x}}\nonumber \\
&\approx&\int\frac{d\mathbf{q}}{\pi^2q}\frac{q^l}{q+\Delta}  e^{i\mathbf{q}\cdot \mathbf{x}}, \nonumber\\
H_{\omega}(\mathbf{x})&=&\int\frac{d\mathbf{q}}{2\pi^2}\lf\frac{1}{q+\Delta-\varepsilon_{12}}-\frac{1}
{q+\Delta+\varepsilon_{12}} \rh e^{i\mathbf{q}\cdot \mathbf{x}}\nonumber \\
&\approx&\varepsilon_{12}\int\frac{d\mathbf{q}}{\pi^2}
\frac{1}{\left(q+\Delta\right)^2}  e^{i\mathbf{q}\cdot \mathbf{x}},\label{omegaaq}
\eee 
where $\Delta=\bar{E}_n-(E_i+E_f)/2 $ and $\varepsilon_{12}=\varepsilon_{1}-\varepsilon_{2}$. 
Here $\varepsilon_{1}$ and $\varepsilon_{2}$ represent 
the energies of the final electrons. Furthermore, contribution of the $p_{1/2}$-wave electrons and term in
 which the nucleon recoil is multiplied by the small $\omega$-term are also neglected.\\
vii)  Since $|\chi U_{ej}g_V'/g_V|\ll|U_{ej}|$, the coupling constant $\chi$ in Hamiltonian (\ref{hamilweakrh}) is neglected. \\
viii) A factorization of phase-space  factors and nuclear matrix elements is achieved by 
the approximation in which electron wave functions $g_{\pm 1}(\varepsilon,r)$,  
$f_{\pm 1}(\varepsilon,r)$ are replaced by their values at the nuclear radius R. 
The notation
\begin{equation}\label{apfns}
g_{\pm 1}(\varepsilon) \equiv g_{\pm 1}(\varepsilon,R), ~~~~~f_{\pm 1}(\varepsilon) \equiv f_{\pm 1}(\varepsilon,R)
\end{equation}
is used. \\

Within above approximations the $0\nu\beta\beta$-decay half-life can be written as
\bee \label{halflifem}
&&\left[T_{1/2}^{0\nu}\right]^{-1} = \frac{\Gamma^{0\nu}}{\ln{2}} = g_A^4\left|M_{GT}\right|^2\left\{C_{mm}\lf
\frac{\left| m_{\beta\beta}  \right|}{m_e}\rh^2 
\right.\nonumber\\ 
&& + C_{m\lambda}\frac{\left| m_{\beta\beta}\right|}{m_e}\left\langle \lambda\right\rangle\cos{\psi_1}
+ C_{m\eta}\frac{\left| m_{\beta\beta}\right|}
{m_e}\left\langle \eta\right\rangle\cos{\psi_2}\nonumber\\  
&& + \left. C_{\lambda\lambda}\left\langle \lambda\right\rangle^2 +C_{\eta\eta}\left\langle \eta\right\rangle^2
+C_{\lambda\eta}\left\langle \lambda\right\rangle\left\langle \eta\right\rangle \cos{(\psi_1-\psi_2)}\phantom{\frac{!}{!}} \right\}.\nonumber\\
\eee
The effective lepton number violating parameters associated with right-handed currents
and their relative phases are given by
\bee\label{effective}
\left\langle \lambda\right\rangle&=&\lambda|\sum_{j=1}^3 U_{ej}T_{ej}^*(g_V'/g_V)|, \nonumber \\
\left\langle \eta\right\rangle&=&\eta|\sum_{j=1}^3 U_{ej}T_{ej}^*|, \nonumber\\
\psi_1&=&\textrm{arg}[(\sum_{j=1}^3 m_j U_{ej}^2 )(\sum_{j=1}^3 U_{ej}T_{ej}^*(g_V'/g_V)^{*}],\nonumber\\
\psi_2&=&\textrm{arg}[(\sum_{j=1}^3 m_j U_{ej}^2 )(\sum_{j=1}^3 U_{ej}T_{ej}^*)^{*}].\nonumber\\
\eee
With help of (\ref{lambdaeta}) and by assuming (\ref{whynot}), $U_0 \simeq V_0$ and $(g_V'/g_V)\simeq 1$
we get  
\begin{eqnarray}
  \left\langle \lambda\right\rangle &\approx& (M_{W_1}/M_{W_2})^2 \frac{m_D}{m_{LNV}} \left|\xi\right|,
  \nonumber \\
  \left\langle \eta\right\rangle    &\approx&- \tan{\zeta} \frac{m_D}{m_{LNV}} \left|\xi\right|,
\end{eqnarray}  
with 
\begin{eqnarray}
|\xi | &=& |c_{23} c_{12}^2 c_{13} s_{13}^2 - c_{12}^3 c_{13}^3 - c_{13}c_{23} c_{12}^2 s_{13}^2\nonumber \\
  && -c_{12}c_{13}\lf c_{13}^2 s_{12}^2+s_{13}^2 \rh | \nonumber\\
  &\simeq& 0.82 
\end{eqnarray}
Here, $c_{ij}\equiv \cos{(\theta_{ij})}$ and $s_{ij}\equiv \sin{(\theta_{ij})}$. $\xi$ was evaluated
by assuming the best fit values for mixing angles $\theta_{12}$, $\theta_{13}$ and $\theta_{23}$
entering the PMNS matrix \cite{lisi}. The experimental upper bound on the mixing angle 
of left and right vector bosons is $\zeta < 0.013$ and if the CP-violating phase in the mixing matrix
for right-handed quarks are small one gets $\zeta < 0.0025$. The flavor and CP-violating processes
of kaons and B-mesons make it possible to deduce lower bound on the mass of the
heavy vector boson $M_{W_2} > 2.9$ TeV \cite{Dev}. In the LRSM there could be additional contributions
to $0\nu\beta\beta$-decay due to the double charged Higgs triplet. However, as pointed in Ref.
\cite{Dev}, in the considered case of type-I seesaw dominance, these contributions can be neglected.

The coefficients $C_I$ (I=$mm$, $m\lambda$, $m\eta$, $\lambda\lambda$, $\eta\eta$ and $\lambda\eta$)
are expressed as combinations of nuclear matrix elements and phase-space factors:
\bee\label{Ccoef}
C_{mm} &=& (1-\chi_F+\chi_T)^2G_{01}, \nonumber\\
C_{m\lambda} &=&-(1-\chi_F+\chi_T)\left[\chi_{2-}G_{03}-\chi_{1+}G_{04}\right], \nonumber\\
C_{m\eta} &=&(1-\chi_F+\chi_T)\nonumber\\
&\times&\left[\chi_{2+}G_{03}-\chi_{1-}G_{04}-\chi_{P}G_{05}+\chi_{R}G_{06}\right], \nonumber\\
C_{\lambda\lambda} &=&\chi_{2-}^2G_{02}+\frac{1}{9}\chi_{1+}^2G_{011}-\frac{2}{9}\chi_{1+}\chi_{2-}G_{010}, \nonumber\\
C_{\eta\eta} &=&\chi_{2+}^2G_{02}+\frac{1}{9}\chi_{1-}^2G_{011}-\frac{2}{9}\chi_{1-}\chi_{2+}G_{010}+\chi_{P}^2G_{08}\nonumber\\
&-&\chi_{P}\chi_{R}G_{07}+\chi_{R}^2G_{09}, \nonumber\\
C_{\lambda\eta} &=&-2[\chi_{2-}\chi_{2+}G_{02}-\frac{1}{9}\lf\chi_{1+}\chi_{2+}+\chi_{2-}\chi_{1-}\rh G_{010}\nonumber\\
&+&\frac{1}{9}\chi_{1+}\chi_{1-}G_{011}].  \\\nonumber
\eee
The explicit form of nuclear matrix elements $M_{GT}$ and their ratios $\chi_I$ is presented
in section (\ref{matrixrh}). The integrated kinematical factors are defined as
\bee \label{phasespacerh}
G_{0k}&=&\frac{G_{\beta}^4m_e^2}{64\pi^5 \ln{2}R^2}\int \delta(\varepsilon_1+\varepsilon_2+M_f-M_i)
 \nonumber \\
 &\times&\left( h_{0k}(\varepsilon_1, \varepsilon_2, R)\cos{\theta}+g_{0k}(\varepsilon_1, \varepsilon_2, R)\right)
 \nonumber \\
 &\times& p_1p_2\varepsilon_1\varepsilon_2d\varepsilon_1d\varepsilon_2d(\cos{\theta}) \nonumber\\
&=& \int_{-1}^{1} \left( \frac{G^\theta_{0k}}{\ln{2}}\cos{\theta} + \frac{G_{0k}}{2}\right) d(\cos{\theta}), 
 \eee
where $k=1,2,\cdots$, 11. $p_1$ and $p_2$ are momenta of electrons and $\theta$ is the angle 
between emitted electrons. The functions $h_{0k}(\varepsilon_1,\varepsilon_2,R)$ 
and $g_{0k}(\varepsilon_1,\varepsilon_2,R)$ are defined in section (\ref{functionstokin}).
These definitions  are independent of the weak axial-vector coupling constant $g_A$.
The quantities $G_{0k}$ are given in units of inverse years. We note that if the standard
wave functions of electron (w.f. A) are assumed $G_{010}=G_{03}$ and $G_{011}=G_{04}$. If in addition
contributions from the induced pseudoscalar term of nucleon current  are neglected the decay rate in
Eq. (\ref{halflifem}) reduces to that given in \cite{Doies}. Quantity $G^{\theta}_{0k}$ is relevant
for the angular correlation between the two electrons. We note that $G^{\theta}_{03}=G^{\theta}_{06}=0.$

%%%%%%%%%%%%%%%%%%%%%%%%%%%%%%%%%%%%%%%%%%%%%%%%%%%%%%%%%%%%%%%%%
%
\subsection{Components due to electron wave functions in the phase-space factors
\label{functionstokin}}
%
%%%%%%%%%%%%%%%%%%%%%%%%%%%%%%%%%%%%%%%%%%%%%%%%%%%%%%%%%%%%%%%

The $s_{1/2}$ and $p_{1/2}$ electron wave functions at nuclear surface   
associated with emission of both electrons enter into the phase-space factors 
through the functions presented below.

For phase-space factors $G^{\theta}_{0k}$ related with the angular distribution
of emitted electrons the quantities $h_{0k}(\epsilon_1,\epsilon_2,R)$ are:
\begin{eqnarray}
h_{01}&=&-4C_{ss}(\varepsilon_1)C_{ss}(\varepsilon_2),\nonumber\\
h_{02}&=&\frac{2\varepsilon_{12}^2}{m_e^2}C_{ss}(\varepsilon_1)C_{ss}(\varepsilon_2),\nonumber\\
h_{03}&=&0,\nonumber\\
h_{04}&=&-\frac{2}{3m_eR} \lf C_{sp}^f(\varepsilon_1)C_{ss}(\varepsilon_2)+C_{sp}^f(\varepsilon_2)C_{ss}(\varepsilon_1)\right.\nonumber\\
&+&\left.C_{sp}^g(\varepsilon_2)C_{ss}(\varepsilon_1)+C_{sp}^g(\varepsilon_1)C_{ss}(\varepsilon_2)\rh , \nonumber\\
h_{05}&=&\frac{4}{m_eR} \lf  C_{sp}^f(\varepsilon_1)C_{ss}(\varepsilon_2)+C_{sp}^f(\varepsilon_2)C_{ss}(\varepsilon_1)\right. \nonumber\\
&+&\left.C_{sp}^g(\varepsilon_2)C_{ss}(\varepsilon_1)+C_{sp}^g(\varepsilon_1)C_{ss}(\varepsilon_2)\rh ,\nonumber\\
h_{06}&=&0,\nonumber\\
h_{07}&=&\frac{-16}{\lf m_eR\rh^2} \lf C_{sp}^f(\varepsilon_1)C_{ss}(\varepsilon_2)+C_{sp}^f(\varepsilon_2)C_{ss}(\varepsilon_1)\right.\nonumber\\
&-&\left.C_{sp}^g(\varepsilon_2)C_{ss}(\varepsilon_1)-C_{sp}^g(\varepsilon_1)C_{ss}(\varepsilon_2) \rh, \nonumber\\
h_{08}&=&\frac{-8}{\lf m_eR\rh^2} \lf C_{sp}^f(\varepsilon_1)C_{sp}^g(\varepsilon_2)+C_{sp}^f(\varepsilon_2)C_{sp}^g(\varepsilon_1)\right.\nonumber\\
&+&\left.C_{ss}(\varepsilon_1)C_{pp}(\varepsilon_2)+C_{ss}(\varepsilon_2)C_{pp}(\varepsilon_1)  \rh , \nonumber\\
h_{09}&=&\frac{32}{\lf m_eR\rh^2} C_{ss}(\varepsilon_1)C_{ss}(\varepsilon_2), \nonumber\\
h_{010}&=&-\frac{9}{2}\tilde{h}_{010}-h_{02},\nonumber\\
h_{011}&=& 9\tilde{h}_{011}+\frac{1}{9}h_{02}+\tilde{h}_{010},
\end{eqnarray}
with
\bee
\tilde{h}_{010}&=&\frac{2\varepsilon_{12}}{3 m_e^2R} \lf  C_{sp}^f(\varepsilon_1)C_{ss}(\varepsilon_2)-C_{sp}^f(\varepsilon_2)C_{ss}(\varepsilon_1) \right.\nonumber \\
&+&\left. C_{sp}^g(\varepsilon_2)C_{ss}(\varepsilon_1)-C_{sp}^g(\varepsilon_1)C_{ss}(\varepsilon_2)\rh, \nonumber \\
\tilde{h}_{011}&=&\frac{-2}{(3 m_eR)^2} \left[ C_{sp}^f(\varepsilon_1)C_{sp}^f(\varepsilon_2)+C_{sp}^g(\varepsilon_2)C_{sp}^g(\varepsilon_1)\right.\nonumber\\
&+&\left.C_{ss}(\varepsilon_1)C_{pp}(\varepsilon_2)+C_{ss}(\varepsilon_2)C_{pp}(\varepsilon_1)\right].
\eee

In addition, the components $g_{0k}(\epsilon_1,\epsilon_2,R)$ of the phase-space factors (\ref{phasespacerh}) are:
\begin{eqnarray}
g_{01}&=&g_{11}=C_{ss}^+(\varepsilon_1)C_{ss}^+(\varepsilon_2),\nonumber\\
g_{02}&=&\frac{\varepsilon_{12}^2}{2 m_e^2}\left(C_{ss}^+(\varepsilon_1)
C_{ss}^+(\varepsilon_2)-C_{ss}^-(\varepsilon_1)C_{ss}^-(\varepsilon_2)\right),\nonumber\\
g_{03}&=&\frac{\varepsilon_{12}}{m_e}\left(C_{ss}^+(\varepsilon_1)C_{ss}^-(\varepsilon_2)
-C_{ss}^+(\varepsilon_2)C_{ss}^-(\varepsilon_1)\right),\nonumber\\
g_{04}&=&\frac{1}{3m_eR} \lf -C_{ss}^-(\varepsilon_1)
C_{sp}^-(\varepsilon_2)-C_{ss}^-(\varepsilon_2)C_{sp}^-(\varepsilon_1)\right.\nonumber \\
&+&\left.C_{ss}^+(\varepsilon_1)C_{sp}^+(\varepsilon_2)+C_{ss}^+(\varepsilon_2)C_{sp}^+(\varepsilon_1)
\rh-g_{03}/9, \nonumber\\
g_{05}&=&\frac{-2}{m_eR} \lf C_{ss}^-(\varepsilon_1)C_{sp}^-(\varepsilon_2)+C_{ss}^-(\varepsilon_2)C_{sp}^-(\varepsilon_1)\right.\nonumber \\
&+&\left.C_{ss}^+(\varepsilon_1)C_{sp}^+(\varepsilon_2)+C_{ss}^+(\varepsilon_2)C_{sp}^+(\varepsilon_1)
\rh, \nonumber\\
g_{06}&=&\frac{4}{m_eR} \lf C_{ss}^+(\varepsilon_1)C_{ss}^-(\varepsilon_2)+C_{ss}^+(\varepsilon_2)C_{ss}^-(\varepsilon_1)\rh,\nonumber\\
g_{07}&=&\frac{-8}{\lf m_eR\rh^2} \lf 
C_{ss}^+(\varepsilon_1)C_{sp}^-(\varepsilon_2)+C_{ss}^+(\varepsilon_2)C_{sp}^-(\varepsilon_1)\right.\nonumber \\
&+& \left.C_{ss}^-(\varepsilon_1)C_{sp}^+(\varepsilon_2)+C_{ss}^-(\varepsilon_2)C_{sp}^+(\varepsilon_1)\rh, \nonumber\\
g_{08}&=&\frac{2}{\lf m_eR\rh^2} \lf- C_{pp}^-(\varepsilon_1)C_{ss}^-(\varepsilon_2)-C_{pp}^-(\varepsilon_2)C_{ss}^-(\varepsilon_1)\right.\nonumber \\
&+&\left.C_{pp}^+(\varepsilon_1)C_{ss}^+(\varepsilon_2)+
C_{pp}^+(\varepsilon_2)C_{ss}^+(\varepsilon_1)\right.\nonumber\\
&+&2C_{sp}^-(\varepsilon_1)C_{sp}^-(\varepsilon_2) +\left.2C_{sp}^+(\varepsilon_1)C_{sp}^+(\varepsilon_2)\rh, \nonumber\\
g_{09}&=&\frac{8}{\lf m_eR\rh^2} \lf C_{ss}^+(\varepsilon_1)C_{ss}^+(\varepsilon_2)+C_{ss}^-(\varepsilon_1)
C_{ss}^-(\varepsilon_2)\rh ,\nonumber\\
g_{010}&=&-\frac{9}{2}\tilde{g}_{010}-g_{02},\nonumber\\
g_{011}&=& 9\tilde{g}_{011}+\frac{1}{9}g_{02}+\tilde{g}_{010},
\end{eqnarray}
with
\bee
\tilde{g}_{010}&=&\frac{\varepsilon_{12}}{3 m_e^2R} \lf 
-C_{ss}^+(\varepsilon_1)C_{sp}^-(\varepsilon_2)+C_{ss}^+(\varepsilon_2)C_{sp}^-(\varepsilon_1)\right.\nonumber \\
&+&\left.C_{ss}^-(\varepsilon_1)C_{sp}^+(\varepsilon_2)-C_{ss}^-(\varepsilon_2)C_{sp}^+(\varepsilon_1)\rh, \nonumber \\
\tilde{g}_{011}&=&\frac{1}{18 m_e^2R^2} \left[ C_{pp}^-(\varepsilon_1)C_{ss}^-(\varepsilon_2)+C_{pp}^-(\varepsilon_2)C_{ss}^-(\varepsilon_1)\right.\nonumber \\
&+&\left.C_{pp}^+(\varepsilon_1)C_{ss}^+(\varepsilon_2)+C_{pp}^+(\varepsilon_2)C_{ss}^+(\varepsilon_1) \right.\nonumber\\
&-&\left.2C_{sp}^-(\varepsilon_1)C_{sp}^-(\varepsilon_2)+2C_{sp}^+(\varepsilon_1)C_{sp}^+(\varepsilon_2) \right].
\eee
Here, $C_{..}$ are combinations of radial components of $s_{1/2}$ and $p_{1/2}$ wave functions,
\begin{eqnarray}\label{cfunctions}
C_{ss}(\varepsilon)&=&g_{-1}(\varepsilon)f_{+1}(\varepsilon) , ~~ C_{pp}(\varepsilon_1)=g_{1}(\varepsilon)f_{-1}(\varepsilon) ,\nonumber\\
C_{sp}^f(\varepsilon)&=&f_{-1}(\varepsilon)f_{+1}(\varepsilon) , ~~ C_{sp}^g(\varepsilon)=g_{-1}(\varepsilon)g_{+1}(\varepsilon),\nonumber\\
C_{ss}^{\pm}(\varepsilon) &=& g^2_{-1}(\varepsilon) \pm f^2_{+1}(\varepsilon), ~~ 
C_{pp}^{\pm}(\varepsilon) = g^2_{+1}(\varepsilon) \pm f^2_{-1}(\varepsilon), \nonumber\\
C_{sp}^{\pm}(\varepsilon) &=&   g_{-1}(\varepsilon) f_{-1}(\varepsilon) \pm g_{+1}(\varepsilon) f_{+1}(\varepsilon).
\end{eqnarray}

%%%%%%%%%%%%%%%%%%%%%%%%%%%%%%%%%%%%%%%%%%%%%%%%%%%%%%%%%%%%%%%%%
%
\subsection{Nuclear matrix elements entering the decay rate \label{matrixrh}}
%
%%%%%%%%%%%%%%%%%%%%%%%%%%%%%%%%%%%%%%%%%%%%%%%%%%%%%%%%%%%%%%%

The expression for the $0\nu\beta\beta$-decay half-life in (\ref{halflifem})
contains matrix element ratios $\chi_I$ and their linear  combinations $\chi_{1\pm }$ and
$\chi_{2\pm }$.  The quantities $\chi_I$ are defined as
\bee
\chi_I=M_I/M_{GT}, 
\eee
where $\textrm{I=F,T, $\omega$F, $\omega$GT, $\omega$T, qF, qGT, qT, R~ and~ P}$
and $M_{GT}$ is the dominant Gamow-Teller matrix element associated with
mechanism due to the left-handed currents. The combinations
$\chi_{1\pm }$ and $\chi_{2\pm }$ are given by
\bee
\chi_{1\pm }
&=&\chi_{qGT}-6\chi_{qT}\pm3\chi_{qF}, \nonumber \\
\chi_{2\pm }&=&\chi_{GT\omega}+\chi_{T\omega}\pm \chi_{F\omega}-\frac{1}{9}\chi_{1\mp}.
\eee
The nuclear matrix elements $M_I$ depend on the exchange potentials $h_I(r)$ through
\bee
&&M_{ F,  GT,  T}=\sum_{rs}\left\langle A_f\left\|h_{  F,   GT, T}(r_-) 
 {\cal{O}}_{F,GT,T}\right\|A_i\right\rangle\nonumber\\
&&M_{\omega F, \omega GT, \omega T}=\sum_{rs}\left\langle A_f\left\|h_{ \omega F,  \omega GT,\omega T}(r_-) 
 {\cal{O}}_{F,GT,T}\right\|A_i\right\rangle\nonumber\\
&&M_{P}=\sum_{rs}i\left\langle A_f \left\|h_{P}(r_-)\tau^+_r\tau^+_s
\frac{(\mathbf{r}_-\times\mathbf{r}_+)}{R^2}\cdot \vec{\sigma}_r\right\|A_i\right\rangle\nonumber \\
&&M_{qF,qGT,qT}=\sum_{rs}\left\langle A_f\left\| h_{qF,qGT,qT}(r_-){\cal{O}}_{F,GT,T}
\right\|A_i\right\rangle\nonumber \\
&&M_R=\sum_{rs}\left\langle A_f\left\|\left[h_{RG}(r_-){\cal{O}}_{GT}+h_{RT}(r_-){\cal{O}}_{T}\right]\right\|A_i\right\rangle\nonumber ~,
\eee
where ${\cal{O}}_{F,GT,T}$ are the familiar operators $1, \vec{\sigma}_1 \cdot \vec{\sigma}_2$ and  
$3(\vec{\sigma}_1 \cdot \hat{r}_{12})(\vec{\sigma}_2 \cdot \hat{r}_{12})$.

The two-nucleon exchange potentials $h_I(r)$ with  = qF, qGT, qT, RG, RT and  P  can be written as
\bee
&&h_I(r) = \frac{2R}{\pi}\int f_I(q,r) \frac{q ~dq}{q +\bar{E}_n -(E_i+E_f)/2},\nonumber\\ 
\eee
where
\bee
f_{GT} &=&  \frac{j_0(q,r)}{g^2_A}\nonumber\\
&\times& \left(g_A^2(q^2)-\frac{g_A(q^2)g_P(q^2)}{m_N}\frac{q^2}{3}+\frac{g_P^2(q^2)}{4m^2_N}\frac{q^4}{3}\right),\nonumber\\
f_{F} &=& \frac{g_V^2(q^2)}{g_A^2}j_0(q r), \nonumber\\
  f_{T} &=&  \frac{j_2(q,r)}{g^2_A}
\left(\frac{g_A(q^2)g_P(q^2)}{m_N}\frac{q^2}{3}-\frac{g_P^2(q^2)}{4m^2_N}\frac{q^4}{3}\right),\nonumber\\
f_{qF}&=&r \frac{g_V^2(q^2)}{g_A^2}j_1(qr)q,\nonumber \\
f_{qGT}&=&\left(\frac{g_A^2(q^2)}{g_A^2} q+3\frac{g_P^2(q^2)}{g_A^2}\frac{q^5}{4m_N^2}\right.\nonumber \\
&+&\left.\frac{g_A(q^2)g_P(q^2)}{g_A^2}\frac{q^3}{m_N}\right)r j_1(qr),\nonumber \\
f_{qT}&=& \frac{r}{3}\left(\left(\frac{g_A^2(q^2)}{g_A^2} 
q-\frac{g_P(q^2)g_A(q^2)}{2g_A^2}\frac{q^3}{m_N}\right)j_1(qr)\right.\nonumber \\
&-&\left.9\frac{g_P^2(q^2)}{2g_A^2}\frac{ q^5}{20m_N^2}\left[ 2j_1(qr)/3-j_3(qr)\right]\right),\nonumber \\
f_{RG}&=&\frac{-R}{3m_N}\left(1+\frac{g_M(q^2)}{g_V(q^2)}\right)\frac{g_A(q^2)g_V(q^2)}{g_A^2}j_0(qr)q^2, \nonumber \\
f_{RT}&=&\frac{-R}{6m_N}\left(1+\frac{g_M(q^2)}{g_V(q^2)}\right)\frac{g_A(q^2)g_V(q^2)}{g_A^2}j_2(qr)q^2, \nonumber \\
f_{P}&=& \frac{R^2}{r}  \frac{g_V(q^2)g_A(q^2)}{g_A^2} j_1(qr)q\nonumber\\ 
\eee  
And the two-nucleon exchange potentials $h_I(r)$ with 
$\textrm{I= $\omega$F, $\omega$GT and  $\omega$T }$ take the form 
\begin{equation}
h_I(r) = \frac{4R}{\pi}\int f_I(q,r) \frac{q^2 ~dq}{\left(q +\bar{E}_n -(E_i+E_f)/2\right)^2}
\end{equation}
where
\bee
f_{\omega F}=f_F,\quad f_{\omega GT}=f_{GT}, \quad f_{\omega T}=f_{T}.
\eee
Here, $\mathbf{r}_+=(\mathbf{r}_r+\mathbf{r}_s)/2$, $\mathbf{r}_-=(\mathbf{r}_r-\mathbf{r}_s)$.  
$\mathbf{r}_{r,s}$ is the coordinate of decaying nucleon and 
$j_i(qr)$ (i=1,2,3) are the spherical Bessel functions. 
It is assumed that $\mathbf{p}_r+\mathbf{p}_r' \simeq  0$, $E_r-E_r' \simeq  0$ and $\mathbf{p}_r-\mathbf{p}_r' \simeq \mathbf{q}$,
where $\mathbf{q}$ is the momentum exchange. The form factors $g_V(q^2)$, $g_A(q^2)$, $g_M(q^2)$ and $g_V(q^2)$ are defined in
\cite{Simdva} and $g_A=1.269$.

If right-handed currents are switched off, all terms  in Eq.(\ref{halflifem}) except
that proportional to $C_1$ vanish. The
connection with the standard $0\nu\beta\beta$-decay formula (\ref{halflifem}) is then
$G_{01} \equiv G^{0\nu}$ and $M_{GT}(1-\chi_F+\chi_T) \equiv M^{0\nu}$.

%%%%%%%%%%%%%%%%%%%%%%%%%%%%%%%%%%%%%%%%%%%%%%%%%%%%%%%%%%%%%%%%%%%%%%%%%%%%%%%
%
\section{Phase-space factors with improved accuracy}
%
%%%%%%%%%%%%%%%%%%%%%%%%%%%%%%%%%%%%%%%%%%%%%%%%%%%%%%%%%%%%%%%%%%%%%%%%%%%%%%%

In Section \ref{Sec:ewf} different ways of the treatment of  radial wave functions
$g_{\pm 1}(\varepsilon,r)$ and $f_{\pm 1}(\varepsilon,r)$ associated 
with emitted electrons in the $s_{1/2}$ and $p_{1/2}$
wave states were presented.  The derivation of the $0\nu\beta\beta$-decay rate was 
accomplished by considering electron wave functions for the point-like nucleus
(wave function B) or an extended one (wave functions A, C and D); that allowed 
to separate phase-space factors and nuclear matrix elements. The
accuracy of the calculation of phase space factors will be discussed next and
the improved phase-space factors associated with mechanisms due to right-handed currents
obtained using screened exact finite-size Coulomb wave functions of emitted electrons
(wave functions D) will be presented.

\begin{table*}[!t]
  \begin{center}
    \caption{Phase-space factors $G_{0k}$ (k=1, $\cdots$, 11) in units $yr^{-1}$ 
      for the $0\nu\beta\beta$-decay of $^{76}$Ge,$^{130}$Te, and $^{150}$Nd. 
      Calculation was performed by assuming different approximations for the radial wave functions
      $g_{\pm 1}$ and $f_{\pm 1}$ of an electron:
    A) The standard approximation of Doi et al. \cite{Doies};
    B) An analytical solution of Dirac equations for a pointlike nucleus is assumed;
    C) An  exact solution of Dirac equations for a uniform charge distribution in
    nucleus is considered; D) The same as the previous case but the electron
    screening is taken into account \cite{iachello}.\label{tabABCD}} 
      \begin{tabular}{lrrrrlrrrrlrrrr}\hline\hline
        &  & $^{76}$Ge & & &  & & $^{130}$Te & & & & & $^{150}$Nd & & \\\cline{2-5}\cline{7-10}\cline{12-15}
  w.f.       &  A~~  &  B~~   & C~~ & D~~ & & A~~  &  B~~   & C~~ & D~~ & & A~~  &  B~~   & C~~ & D~~\\ \hline
$ G_{01} .10^{14}$&$0.261$ &$0.244$&$0.240$&$0.237$ & &$ 1.807$&$ 1.535$&$ 1.453$&$ 1.425$ & &$ 8.827$ &$  6.986$&$ 6.432$&$ 6.316$  \\
$ G_{02} .10^{14}$&$0.428$ &$0.404$&$0.397$&$0.391$ & &$ 4.683$&$ 4.064$&$ 3.851$&$ 3.761$ & &$40.190$ &$ 32.401$&$29.869$&$29.187$  \\
$ G_{03} .10^{15}$&$1.478$ &$1.340$&$1.316$&$1.305$ & &$12.237$&$ 9.566$&$ 9.065$&$ 8.967$ & &$70.032$ &$ 49.465$&$45.593$&$45.130$  \\
$ G_{04} .10^{15}$&$0.501$ &$0.489$&$0.477$&$0.470$ & &$ 3.625$&$ 3.315$&$ 3.086$&$ 3.021$ & &$18.343$ &$ 16.000$&$14.348$&$14.066$  \\
$ G_{05} .10^{13}$&$0.791$ &$0.727$&$0.572$&$0.566$ & &$ 6.390$&$ 5.185$&$ 3.842$&$ 3.790$ & &$28.537$ &$ 21.183$&$15.061$&$14.873$  \\
$ G_{06} .10^{12}$&$0.605$ &$0.547$&$0.536$&$0.531$ & &$ 3.091$&$ 2.398$&$ 2.258$&$ 2.227$ & &$11.922$ &$  8.323$&$ 7.591$&$ 7.497$  \\
$ G_{07} .10^{10}$&$0.365$ &$0.345$&$0.274$&$0.270$ & &$ 2.713$&$ 2.383$&$ 1.788$&$ 1.755$ & &$13.625$ &$ 11.362$&$ 8.233$&$ 8.085$  \\
$ G_{08} .10^{11}$&$0.245$ &$0.236$&$0.151$&$0.149$ & &$ 2.877$&$ 2.653$&$ 1.579$&$ 1.549$ & &$16.833$ &$ 14.996$&$ 8.564$&$ 8.405$  \\
$ G_{09} .10^{10}$&$1.360$ &$1.263$&$1.238$&$1.223$ & &$ 6.398$&$ 5.354$&$ 5.063$&$ 4.972$ & &$27.582$ &$ 21.530$&$19.799$&$19.454$  \\
$ G_{010}.10^{15}$&$1.478$ &$1.531$&$1.423$&$1.410$ & &$12.237$&$14.602$&$11.616$&$11.455$ & &$70.032$ &$105.415$&$72.249$&$71.154$  \\
$ G_{011}.10^{15}$&$0.501$ &$0.500$&$0.484$&$0.476$ & &$ 3.625$&$ 3.564$&$ 3.220$&$ 3.148$ & &$18.343$ &$ 18.334$&$15.376$&$15.055$  \\   
\hline\hline
      \end{tabular}
  \end{center}
\end{table*}

\begin{figure}[!t]
  \includegraphics[width=1.05\columnwidth]{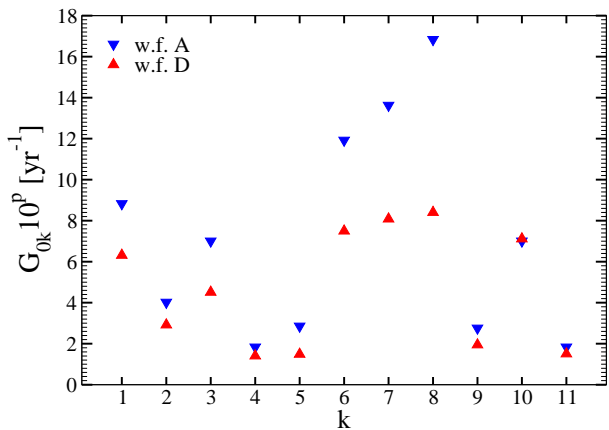}
  \caption{The phase factors $G_{0k}$ (k=1, $\cdots$, 11) in units of $yr^{-1}$ 
    for the $0\nu\beta\beta$-decay of $^{150}$Nd. Results are presented
    for approximate electron wave functions (type A \cite{Doies}) and exact Dirac wave
    functions with finite nuclear size and electron screening (type D \cite{iachello}).
    The exponents p for k=1, $\cdots$, 11 are 14, 14, 15, 15, 13, 12, 10, 11, 10, 15,
    15, respectively.   } 
\label{fig.FermiFun150Nd}
\end{figure}

A numerical computation of all 11 phase-space factors entering the $0\nu\beta\beta$-decay rate
was performed by using previously described 4 types of wave functions  (A, B, C and D) for a sample of
3  isotopes ($^{76}$Ge, $^{130}$Te and $^{150}$Nd). Results are presented in Table 
\ref{tabABCD}. We see that by using the standard treatment of electron wave functions,
corresponding to leading finite-size Coulomb corrections (wave functions A), a significant difference
with results of the other three approaches appear, especially for nuclei with large nuclear charge Z.
Surprisingly, results obtained with wave functions B, corresponding to
an analytical solution of Dirac equations for a pointlike nucleus, better agree with
results corresponding to wave functions C and D (exact solution of Dirac equations for a uniform
nuclear charge distribution with the radius R) than those obtained
by the standard treatment of wave functions (wave functions A). It indicates that 
the exact treatment of the Coulomb field plays more important role than the position of decaying nucleon in the nucleus.
From the Table \ref{tabABCD} it is apparent that effect of the screening of atomic electrons
on the wave functions of emitted electrons does not play an important role. 

The phase-factors differ among themselves significantly in magnitude. This fact is manifested in
Fig. \ref{fig.FermiFun150Nd}. One can see that phase-factors obtained with standard wave functions (w.f. A)
are always larger than those with phase factors calculated with the more advanced wave functions.
Exact treatment reduces the value of all studied phase-space factors.

The phase-space factors (see Eq. (\ref{phasespacerh})) contain products 
of $g_{\pm 1}(\varepsilon )$ and $f_{\pm 1}(\varepsilon)$, which
were evaluated at $r = R$ from radial wave functions $g_{\pm 1}(\varepsilon, r)$ and $f_{\pm 1}(\varepsilon, r)$
(see Eq. (\ref{apfns})). Thus, it is assumed that the $\beta$-decay of both
nucleons happens at the nuclear surface. This procedure can be generalized
with help of normalized to unity distribution function $D(r_1)$ as follows:
\begin{eqnarray}
g_{\pm1}(\varepsilon,R) &=& \int_0^{\infty}  g_{\pm1}(\varepsilon, r_1) D(r_1) dr_1, \nonumber\\
f_{\pm1}(\varepsilon,R) &=& \int_0^{\infty}  f_{\pm1}(\varepsilon, r_1) D(r_1) dr_1
\label{Eq.Point}
\end{eqnarray}
where in this particular case $D(r_1)=\delta(r_1-R)$.

\begin{table}[!t]
  \begin{center}
      \caption{The phase-space factor G$_{01}$ 
      in units $yr^{-1}$ for the $0\nu\beta\beta$-decay of $^{76}$Ge, $^{130}$Te, and $^{136}$Xe. 
      Results are presented for exact Dirac wave functions with finite nuclear size
      and electron screening at nuclear radius R (Exact, approximation scheme D) and those averaged 
      over the distribution deduced from the analysis of the dominant nuclear matrix element
      (Exact and averaged - see Eqs.(\ref{Eq.Point})).  \label{Tab.Priemer}}
      \begin{tabular}{lrrr}\hline\hline
        & \multicolumn{3}{c}{   G$_{01}.10^{14}$ }\\ \cline{2-4}
     & $^{76}$Ge & $^{130}$Te & $^{136}$Xe \\ \hline
       Exact         & $0.23681$ & $1.42547$ & $1.46187$ \\
  Exact and averaged & $0.23987$ & $1.47396$ & $1.52851$ \\   
\hline\hline
      \end{tabular}
  \end{center}
\end{table}

\begin{center}
  \begin{figure}[!t]
  \includegraphics[width=1.05\columnwidth]{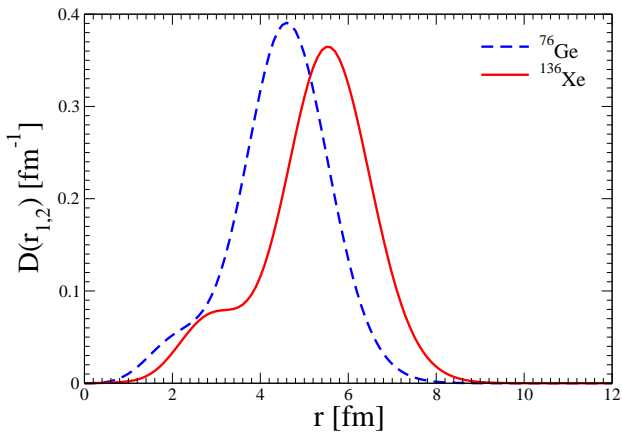}
  \caption{The normalized $r_1$ ($r_2$) dependence of $M^{0\nu}$ for
    $^{76}$Ge and $^{136}Xe$. $r_1$ and $r_2$ are absolute values of a position vector
  of $\beta$-decaying nucleon in a nucleus.} 
\label{fig:rdep}
\end{figure}
\end{center}

In Fig. \ref{fig:rdep} the distribution function $D(r_1)$ (or equivalently $D(r_2)$) 
is shown corresponding to the nuclear matrix element $M^{0\nu}$ (associated with the
$m_{\beta\beta}$ mechanism) of the $0\nu\beta\beta$-decay of $^{76}$Ge and $^{136}$Xe and calculated within
quasiparticle random phase approximation with restoration of isospin symmetry \cite{vadimp}.
We see that $\beta$-decay of both nucleons happens mostly in the vicinity of nuclear surface. 
In Table \ref{Tab.Priemer} the phase space factor G$_{01}$ calculated with help of 
$D(r_1)=\delta(r_1-R)$ and $D(r_1)$ deduced from calculated nuclear matrix elements
of  $^{76}$Ge, $^{130}$Te and $^{136}$Xe are compared. We see that the corresponding
effect is very small for $^{76}$Ge and is only about 4-5 \% in the case of $^{130}$Te and $^{136}$Xe.

The improved phase-space factors $G_{0j}$ (j=1, $\cdots$, 11) in units $yr^{-1}$ associated
with left and right-handed mechanism of the $0\nu\beta\beta$-decay for nuclei
of experimental interest are presented in Table \ref{tab:phas}. They were 
obtained using screened exact finite-size Coulomb wave functions for $s_{1/2}$ and $p_{1/2}$
states of electron (wave functions D). In Table \ref{tab:phasT} the
phase space factors $G^{\theta}_{0j}$ associated with angular distribution of emitted electrons
are presented.

\begin{table*}[!t]
  \begin{center}
    \caption{Phase-space factors $G_{0j}$ (j=1, $\cdots$, 11) in units $yr^{-1}$ 
obtained using screened exact finite-size Coulomb wave functions for $s_{1/2}$ and $p_{1/2}$
states of electron (wave functions D). 
The Q values were taken from experiment when available, or from tables of recommended value \cite{iachello}.
$G_{01}$ is associated with the mechanism generated just by $m_{\beta\beta}$. In the  case 
of dominance of the $\langle \lambda \rangle$ ($\langle \eta \rangle$) mechanism the decay rate
includes phase factors  $G_{02}$, $G_{010}$ and $G_{011}$ ($G_{02}$, $G_{07}$, $G_{08}$, $G_{09}$,
$G_{010}$ and $G_{011}$). The remaining
phase factors are due to interference of these mechanisms (see Eq. (\ref{phasespacerh})).
\label{tab:phas}}
\centering 
\renewcommand{\arraystretch}{1.1}  
      \begin{tabular}{lrrrrrrrrrrr}\hline\hline
          & $^{48}$Ca & $^{76}$Ge & $^{82}$Se & $^{96}$Zr & $^{100}$Mo & $^{110}$Pd & $^{116}$Cd & $^{124}$Sn & $^{130}$Te & $^{136}$Xe & $^{150}$Nd \\ \hline
Q$_{\beta\beta}$ [MeV]& $4.27226$ & $2.03904$ & $2.99512$ & $3.35037$ & $3.03440$ & $2.01785$ & $2.8135$ & $2.28697$ & $2.52697$ & $2.45783$ & $3.37138$ \\
  & & & & & & & & & & & \\	      
 G$_{01}.10^{14}$ & $ 2.483$ & $0.237$ & $1.018$ & $ 2.062$ & $ 1.595$ & $0.483$ & $ 1.673$ & $0.906$ & $1.425$ & $1.462$ & $ 6.316$ \\
 G$_{02}.10^{14}$ & $16.229$ & $0.391$ & $3.529$ & $ 8.959$ & $ 5.787$ & $0.814$ & $ 5.349$ & $1.967$ & $3.761$ & $3.679$ & $29.187$ \\
 G$_{03}.10^{15}$ & $18.907$ & $1.305$ & $6.913$ & $14.777$ & $10.974$ & $2.672$ & $11.128$ & $5.403$ & $8.967$ & $9.047$ & $45.130$ \\
 G$_{04}.10^{15}$ & $ 5.327$ & $0.470$ & $2.141$ & $ 4.429$ & $ 3.400$ & $0.978$ & $ 3.569$ & $1.886$ & $3.021$ & $3.099$ & $14.066$ \\
 G$_{05}.10^{13}$ & $ 3.007$ & $0.566$ & $2.004$ & $ 4.120$ & $ 3.484$ & $1.400$ & $ 4.060$ & $2.517$ & $3.790$ & $4.015$ & $14.873$ \\
 G$_{06}.10^{12}$ & $ 3.984$ & $0.531$ & $1.733$ & $ 3.043$ & $ 2.478$ & $0.934$ & $ 2.563$ & $1.543$ & $2.227$ & $2.275$ & $ 7.497$ \\
 G$_{07}.10^{10}$ & $ 2.682$ & $0.270$ & $1.163$ & $ 2.459$ & $ 1.927$ & $0.599$ & $ 2.062$ & $1.113$ & $1.755$ & $1.812$ & $ 8.085$ \\
 G$_{08}.10^{11}$ & $ 1.109$ & $0.149$ & $0.708$ & $ 1.755$ & $ 1.420$ & $0.462$ & $ 1.703$ & $0.939$ & $1.549$ & $1.657$ & $ 8.405$ \\
 G$_{09}.10^{10}$ & $16.246$ & $1.223$ & $4.779$ & $ 8.619$ & $ 6.540$ & $1.939$ & $ 6.243$ & $3.301$ & $4.972$ & $4.956$ & $19.454$ \\
G$_{010}.10^{14}$ & $ 2.116$ & $0.141$ & $0.801$ & $ 1.855$ & $ 1.359$ & $0.309$ & $ 1.418$ & $0.660$ & $1.146$ & $1.165$ & $ 7.115$ \\
G$_{011}.10^{15}$ & $ 5.376$ & $0.476$ & $2.183$ & $ 4.557$ & $ 3.502$ & $1.010$ & $ 3.704$ & $1.955$ & $3.148$ & $3.238$ & $15.055$ \\
   \hline\hline
      \end{tabular}
  \end{center}
\end{table*}

  \begin{table*}[!t]
  \begin{center}
    \caption{Phase space factors $G^{\theta}_{0j}$ associated with angular distribution of emitted electrons 
      (see Eq. (\ref{phasespacerh})) in units $yr^{-1}$ obtained using screened exact finite-size Coulomb
      wave functions for $s_{1/2}$ and $p_{1/2}$ states of electron (wave functions D). The Q-values of 
      Tab.(\ref{tab:phas}) are assumed.
\label{tab:phasT}}
\centering 
\renewcommand{\arraystretch}{1.1}  
      \begin{tabular}{rrrrrrrrrrrr}\hline\hline
          & $^{48}$Ca & $^{76}$Ge & $^{82}$Se & $^{96}$Zr & $^{100}$Mo & $^{110}$Pd & $^{116}$Cd & $^{124}$Sn & $^{130}$Te & $^{136}$Xe & $^{150}$Nd \\ \hline
  & & & & & & & & & & & \\	      
 -G$^{\theta}_{01}.10^{15}$  &   8.010  & 0.679 &  3.141 &  6.484  & 4.951 &  1.397 &  5.153 &  2.699 &  4.328 &  4.426 & 20.101  \\
  G$^{\theta}_{02}.10^{14}$  &   5.144  & 0.113 &  1.075 &  2.769  & 1.770 &  0.239 &  1.629 &  0.587 &  1.137 &  1.111 &  9.138  \\
 -G$^{\theta}_{04}.10^{15}$  &   1.786  & 0.152 &  0.703 &  1.456  & 1.112 &  0.314 &  1.161 &  0.608 &  0.977 &  1.000 &  4.566  \\
  G$^{\theta}_{05}.10^{15}$  &  10.714  & 0.910 &  4.219 &  8.734  & 6.674 &  1.884 &  6.965 &  3.650 &  5.862 &  6.002 & 27.397  \\
  G$^{\theta}_{07}.10^{11}$  &   8.458  & 0.711 &  3.422 &  7.431  & 5.706 &  1.589 &  6.026 &  3.089 &  5.016 &  5.155 & 24.824  \\
  G$^{\theta}_{08}.10^{12}$  &   3.553  & 0.402 &  2.121 &  5.383  & 4.271 &  1.251 &  5.054 &  2.651 &  4.498 &  4.787 & 26.100  \\
  G$^{\theta}_{09}.10^{10}$  &   5.024  & 0.313 &  1.379 &  2.562  & 1.904 &  0.504 &  1.795 &  0.899 &  1.397 &  1.387 &  5.899  \\
 G$^{\theta}_{010}.10^{15}$  &   0.695  & 0.028 &  0.317 &  1.118  & 0.764 &  0.114 &  0.884 &  0.334 &  0.706 &  0.741 &  7.816  \\
-G$^{\theta}_{011}.10^{15}$  &   1.790  & 0.152 &  0.707 &  1.466  & 1.120 &  0.317 &  1.172 &  0.615 &  0.988 &  1.012 &  4.594  \\
\hline\hline
      \end{tabular}
  \end{center}
\end{table*}

%%%%%%%%%%%%%%%%%%%%%%%%%%%%%%%%%%%%%%%%%%%%%%%%%%%%%%%%%%%%%%%%%%%%%%%%%%%%%%%%%%%%%%%%%%%%%%%%%%%%
%
\section{Constraints on the effective total lepton number violating parameters}
%
%%%%%%%%%%%%%%%%%%%%%%%%%%%%%%%%%%%%%%%%%%%%%%%%%%%%%%%%%%%%%%%%%%%%%%%%%%%%%%%%%%%%%%%%%%%%%%%%%%%%

Experimental $0\nu\beta\beta$-decay half-life limits may be used, in combination with the
formula  (\ref{halflifem}), to constrain the effective Majorana neutrino mass $m_{\beta\beta}$
and the effective coupling constants $\langle\lambda\rangle$  and $\langle\eta\rangle$
of the right-handed currents.  This can be done provided the
values of phase-space factors and nuclear matrix elements are available. 
We shall use the quasiparticle random phase approximation (QRPA) \cite{muto89}
and interacting shell-model (ISM) \cite{cau96} matrix elements for such 
analysis. Their values are presented in Table \ref{tab.MatElem}. 
In the case of ISM the magnitude of matrix elements $M_{GT}$ calculated in \cite{cau08} are assumed.
We note that these matrix elements were evaluated when the contribution
from the induced pseudoscalar term of hadron current was not taken into account. In analysis below
the case of CP conservation ($\psi_1=\psi_2=0$) is assumed. 

\begin{table}[!t]
  \begin{center}
    \caption{Nuclear matrix elements and their ratios. The quasiparticle random phase approximation (QRPA)
     and interacting shell-model (ISM) matrix elements are from \cite{muto89} and  \cite{cau96}, respectively.
     In the case of ISM matrix elements $M_{GT}$ calculated in \cite{cau08} is used.
      \label{tab.MatElem}}
 \begin{tabular}{lccccccc}\hline\hline
		\renewcommand{\arraystretch}{1.3}  
  & &   \multicolumn{2}{c}{ $^{76}$Ge }  & & \multicolumn{2}{c}{ $^{136}$Xe } & \\ \cline{3-4} \cline{6-7}
  & & ISM & QRPA & & ISM & QRPA & \\ \hline
$M_{GT}$         & &   2.350   &  3.014   & &  1.770  &  1.120  & \\
$\chi_F$        & &  -0.106   &  -0.389  & &  -0.151 &  -0.412 & \\
$\chi_{1+}$      & &  0.686    &  0.811   & &  0.782  &  1.969 & \\
$\chi_{1-}  $    & &  1.340    &  2.917   & &  1.784  &   4.052 & \\
$\chi_{2+}  $    & &  0.633    &  0.302   & &  0.556  &   0.229 &\\
$\chi_{2-}  $    & &  0.912    &  1.216   & &  0.965  &   1.195 & \\
$\chi_{R}  $     & &  0.684    &  1.192   & &  0.955  &   1.958 & \\
$\chi_{P}  $     & & -0.544    &  -0.176  & &  0.256  &   -0.321 & \\
\hline\hline
      \end{tabular}
  \end{center}
\end{table}

\begin{center}
  \begin{figure}[!t]
    \includegraphics[width=1.05\columnwidth]{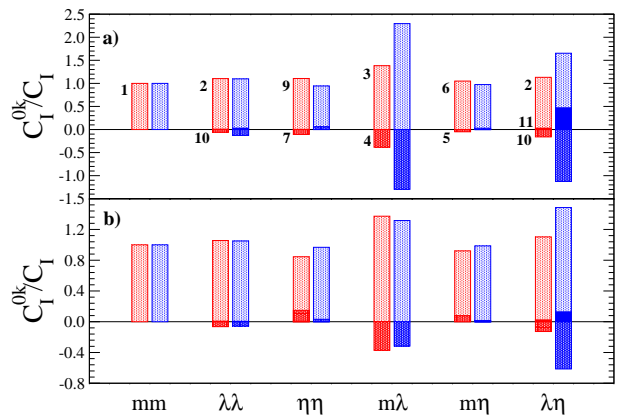}
  \caption{The decomposition of coefficients 
    $C_I$ (I=$mm$, $m\lambda$, $m\eta$, $\lambda\lambda$, $\eta\eta$ and $\lambda\eta$,
    see Eqs. (\ref{halflifem}) and (\ref{Ccoef}) ) on partial contributions $C^{0k}_I$ associated with
    phase-spase factors $G_{0k}$ (k=1, $\cdots$,11). The symbols standing for index I are shown
    on the x-axis. The partial contributions are identified by index k, which value is shown by
    the corresponding slope.  Ratios $C^{0k}_I/C_I$ calculated with the ISM and QRPA matrix elements
    are displayed with left (in red) and right (in blue) slopes, respectively. Results for $^{76}$Ge
    and $^{136}$Xe are presented in the lower  b) and upper a) panels, respectively.
  }
\label{fig:cfac}
\end{figure}
\end{center}

Different contributions to the  $0\nu\beta\beta$-decay rate (\ref{halflifem})  are associated
with different products of effective lepton number violating parameters $m_{\beta\beta}$,
$\langle\lambda\rangle$  and $\langle\eta\rangle$, which values are unknown.
The importance of these contributions depends also on the value coefficients 
$C_I$ (I=$mm$, $m\lambda$, $m\eta$, $\lambda\lambda$, $\eta\eta$ and $\lambda\eta$,
which can be calculated. The quantity is a superposition of contributions $C^{0k}_I$ associated with
phase-spase factors $G_{0k}$ (k=1, $\cdots$,11). In Fig. \ref{fig:cfac} we show ratios
$C^{0k}_I/C_I$ for the $0\nu\beta\beta$-decay $^{76}$Ge and $^{136}$Xe and both sets
of nuclear matrix elements. We note that coefficients $C_{mm}$, $C_{\lambda\lambda}$, $C_{\eta\eta}$,
and $C_{m\eta}$ are dominated by a single contribution associated with a different phase-factor.
In the case of $C_{m\lambda}$ and  $C_{\lambda\eta}$ there is a competition of mostly two contributions.

\begin{figure}[!t]
  \includegraphics[width=1.05\columnwidth]{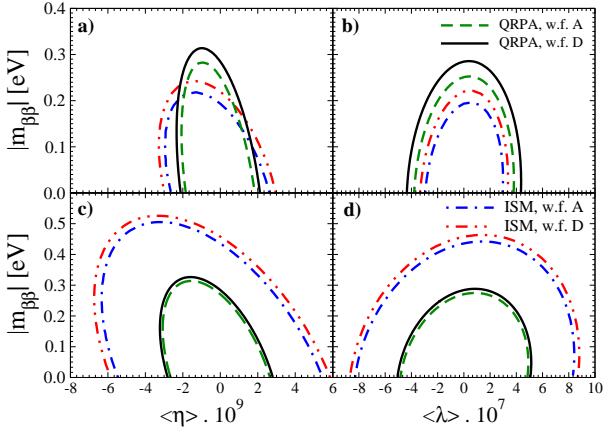}
  \caption{
    Limits on the effective neutrino mass $m_{\beta\beta}$ and right handed parameters 
    $\eta$  (left panels, $\langle\lambda\rangle =0$) and $\lambda$ (right panels, $\langle\eta\rangle =0$)
    implied by the constraints on the $0\nu\beta\beta$-decay of $^{76}$Ge (lower panels,
    $T^{0\nu}_{1/2} \ge 3.0\times 10^{25}$ \cite{gelimit}) and
    $^{136}$Xe (upper panels, $T^{0\nu}_{1/2} \ge 3.4\times 10^{25}$  \cite{xelimit}). To derive the bounds, the
    values of nuclear matrix elements calculated within the ISM \cite{cau96} and the QRPA \cite{muto89}
    are used. Results are presented
    for approximate electron wave functions (type A) and exact Dirac wave
    functions with finite nuclear size and electron screening (type D).
    Ellipsoids show the boundaries of the allowed domains.} 
\label{fig.restriction}
\end{figure}

Using these nuclear matrix elements  (Table \ref{tab.MatElem}) and the
phase-space factors calculated here (see Table \ref{tab:phas})
we can deduce from the experimental data 
$T^{0\nu}_{1/2} \ge 3.0\times 10^{25}$ for $^{76}$Ge  decay \cite{gelimit}
and $T^{0\nu}_{1/2} \ge 3.4\times 10^{25}$  for $^{136}$Xe decay \cite{xelimit}
(we use here the combined limit from the EXO and KamLAND-Zen experiments)
the constraints on the effective right-handed current couplings
$\langle\lambda\rangle$, $\langle\eta\rangle$  and the effective Majorana
neutrino mass $m_{\beta\beta}$ listed in Table \ref{tab.constraint}.
The constraints in Table \ref{tab.constraint} are of a similar magnitude as
those in Table I of Ref. \cite{Dev}. However, they are based now on the exact
treatment of the phase-space factors as well as on the more complete account of
nuclear matrix elements.
Fig. \ref{fig.restriction} shows the allowed regions for
$m_{\beta\beta}$ and $\langle\lambda\rangle$ ($\langle\eta\rangle$) for 
$\langle\eta\rangle =0$ ($\langle\lambda\rangle =0$). Results are presented
for the two sets of nuclear matrix elements (ISM \cite{cau96,cau08} and QRPA
\cite{muto89}) and the standard (w.f. A) and improved (w.f. D)
description of electron wave functions. Note that limits on
lepton number violating parameters are softened a little 
when other lepton number violating parameters
have non-vanishing values at the same time in comparison with the case
when only a single parameter is non-zero. By assuming 
$\zeta$ = 0.013 and  0.0025 mentioned earlier and the current limit
$\langle\eta\rangle \le 2.98\times 10^{-9}$ ($^{136}$Xe, ISM, w.f. D)
we end up with $m_D/m_{LNV} = 2.8\times 10^{-7}$ and $1.5\times 10^{-6}$, respectively.
For $M_{W_2} = 2.9$ TeV and $\langle\lambda\rangle \le 3.34\times 10^{-7}$
($^{136}$Xe, ISM, w.f. D) we get $m_D/m_{LNV} = 5.0\times 10^{-6}$.
Thus, from the more stringent limits on $\langle\eta\rangle$ we obtain
$m_{LNV}$/TeV = 0.3 - 2 $m_D$/MeV, in agreement with the assumption that the basic
scale of LRSM is O(TeV). It is therefore obvious that already the present
limits of $0\nu\beta\beta$-decay half-lives can be used to constrain meaningfully
the allowed parameter space of LRSM, and that mechanism associated with
right-handed currents can compete with the one based on $m_{\beta\beta}$ that
is so often used.

\begin{table}[!t]
  \begin{center}
    \caption{Upper bounds on the effective neutrino mass $m_{\beta\beta}$
      and parameters $\langle\eta\rangle$ and $\langle\lambda\rangle$ associated with right-handed
      currents mechanisms 
    imposed by the constraints on the $0\nu\beta\beta$-decay of $^{76}$Ge (upper panels,
    $T^{0\nu}_{1/2} \ge 3.0\times 10^{25}$ \cite{gelimit}) and
    $^{136}$Xe ($T^{0\nu}_{1/2} \ge 3.4\times 10^{25}$  \cite{xelimit}).
    Nuclear matrix elements of interacting shell-model (ISM) \cite{cau96} ($M_{GT}$ is from \cite{cau08})
    and quasiparticle random phase approximations (QRPA) \cite{muto89} are used in analysis.
    CP conservation is assumed ($\psi_1=\psi_2=0$). The standard electron wave functions (w.f. A) \cite{Doies}
    and screened exact finite-size Coulomb  wave functions (w.f. D) are considered.\label{tab.constraint}}
\begin{tabular}{lccccccc}\hline\hline
 		\renewcommand{\arraystretch}{1.3}  
  & &   \multicolumn{2}{c}{ $^{76}$Ge }  & & \multicolumn{2}{c}{ $^{136}$Xe } & \\ \cline{3-4} \cline{6-7}
 w.f.	              & &   A  & D & &  A  & D &  \\ \hline
 & \multicolumn{7}{c}{ QRPA} \\
$|m_{\beta\beta}|$   [eV]        & &  0.321  &  0.333  & & 0.285 & 0.315 & \\
 $|m_{\beta\beta}|$ [eV] (for $\langle\eta\rangle=\langle\lambda\rangle=0$) & & 0.271   &  0.284  & & 0.251 & 0.285 & \\
 $\langle\eta\rangle\times 10^{-9}$         & & 3.093   &  3.239  & & 2.077 & 2.337 & \\
$\langle\eta\rangle\times 10^{-9}$   [eV] (for $\langle\eta\rangle=\langle\lambda\rangle=0$)  & & 2.652   &  2.807  & &1.840 & 2.118 & \\
$\langle\lambda\rangle \times 10^{-7}  $   & & 4.943   &  5.163  & & 3.822 & 4.370 & \\
$\langle\lambda\rangle\times 10^{-7}$   [eV] (for $\langle\eta\rangle=\langle\lambda\rangle=0$)  & & 4.841   &  5.068  & & 3.792 & 4.349 & \\
 & \multicolumn{7}{c}{ ISM} \\
$|m_{\beta\beta}|$   [eV]        & & 0.515   &  0.535  & & 0.222 & 0.245 & \\
$|m_{\beta\beta}|$   [eV] (for $\langle\eta\rangle=\langle\lambda\rangle=0$)  & & 0.436   &  0.458  & & 0.194 & 0.220 & \\
$\langle\eta\rangle\times 10^{-9}$         & & 6.370   &  6.760  & & 2.975 & 3.291 & \\
$\langle\eta\rangle\times 10^{-9}$   [eV] (for $\langle\eta\rangle=\langle\lambda\rangle=0$)  & & 5.464   &  5.863  & & 2.628 & 2.976 & \\
$\langle\lambda\rangle \times 10^{-7}  $   & & 8.462   &  8.841  & & 3.000 & 3.378 & \\
$\langle\lambda\rangle\times 10^{-7}$   [eV] (for $\langle\eta\rangle=\langle\lambda\rangle=0$)  & & 8.304   &  8.694 & & 2.949 & 3.336 & \\
\hline\hline
      \end{tabular}
  \end{center}
\end{table}

%%%%%%%%%%%%%%%%%%%%%%%%%%%%%%%%%%%%%%%%%%%%%%%%%%%%%%%%%%%%%%%%%%%%%%%%%%%%%%%%%%%%%%%%%%%%%%%%%%%%
%
\section{ Differential decay rates for limiting cases}
%
%%%%%%%%%%%%%%%%%%%%%%%%%%%%%%%%%%%%%%%%%%%%%%%%%%%%%%%%%%%%%%%%%%%%%%%%%%%%%%%%%%%%%%%%%%%%%%%%%%%%

It is of interest to consider the angular correlations of the emitted electrons and
the single electron energy spectrum for the three limiting cases of lepton number violating mechanism
since  with sufficient experimental accuracy one could distinguish between decays due to coupling to
the left-handed and right-handed hadronic currents. It is assumed that
some future $0\nu\beta\beta$-decay experiments, e.g. the SuperNEMO \cite{Super} or NEXT \cite{Next},
will have a unique potential to measure the electron tracks and thus to observe the decay electron angular
correlations and individual electron energy spectra.

\begin{center}
  \begin{figure}[!t]
  \includegraphics[width=1.0\columnwidth]{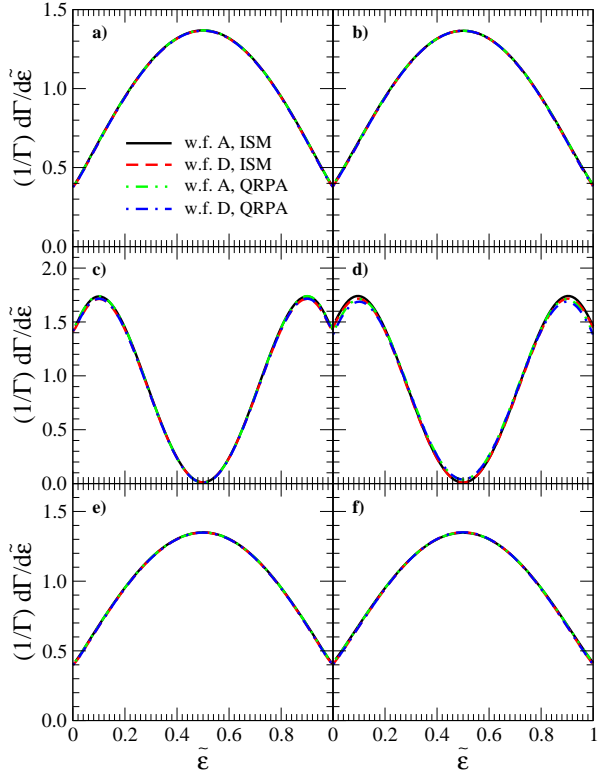}
  \caption{The single electron differential decay rate normalized
    to the total decay rate vs. the electron energy ${\tilde \varepsilon}$
    (${\tilde \varepsilon}=(\varepsilon - m_e)/Q_{\beta\beta}$) 
    for $0\nu\beta\beta$-decay of $^{76}$Ge (left panels a), c) and e))
    and $^{136}$Xe (right panels b), d) and f)). Results are presented
    for $|m_{\beta\beta}|^2$ (panels a) and b)); $\langle \lambda \rangle = \langle \eta \rangle = 0$),
    $|\langle \lambda \rangle|^2$ (panels c) and d)); $m_{\beta\beta} = \langle \eta \rangle = 0$),
    $|\langle \eta \rangle|^2$ (panels e) and f)); $m_{\beta\beta} = \langle \lambda \rangle = 0$)
    terms. An approximate electron wave functions (w.f. A) and exact Dirac wave
    functions with finite nuclear size and electron screening (w.f. D) are considered.
    Nuclear matrix elements calculated within the ISM \cite{cau96} and the QRPA \cite{muto89}
    are used in calculations.} 
\label{fig:endis}
\end{figure}
\end{center}

The differential rate for the $0^+\rightarrow 0^+$ $0\nu\beta\beta$-decay 
with the energy of one of the emitted electrons
$\tilde{\varepsilon}_1$ ( $\tilde{\varepsilon}_1$ is the kinetic energy fraction with 
respect to $Q_{\beta\beta}$ of one electron, i.e. $\varepsilon_1=\tilde{\varepsilon}_1 Q_{\beta\beta} +m_e$
and $\varepsilon_2=Q_{\beta\beta}+2m_e-\varepsilon_1$) and the 
angular distribution with the angle $\theta$ between the two electrons for three limiting cases 
can be written as follows:\\
i) Case $m_{\beta\beta}\ne 0$  ($\left\langle \lambda\right\rangle=0$ and 
  $\left\langle \eta\right\rangle=0$),
  \begin{eqnarray}
d \Gamma
&=& g_A^4 \left|M_{GT}\right|^2
\lf \frac{\left| m_{\beta\beta}\right|}{m_e}\rh^2 dC_{mm}.
  \end{eqnarray}
\\
ii) Case $\left\langle \lambda\right\rangle \ne0$ ($m_{\beta\beta}=0$ 
  and $\left\langle \eta\right\rangle=0$), 
\be
d \Gamma
= g_A^4 \left|M_{GT}\right|^2
\left\langle \lambda\right\rangle^2 dC_{\lambda\lambda}.
\ee
\\
iii) Case $\left\langle \eta\right\rangle \ne 0$ ($m_{\beta\beta}=0$ and
  $\left\langle \lambda\right\rangle =0$),
\be
d \Gamma
= g_A^4 \left|M_{GT}\right|^2
 \left\langle \eta\right\rangle^2 dC_{\eta\eta},
\ee
where 
\bee
d C_{mm} &=&(1-\chi_F)^2 d{\cal G}_{01}, \nonumber\\
dC_{\lambda\lambda} &=& \chi_{2-}^2d{\cal G}_{02}+\frac{1}{9}\chi_{1+}^2d{\cal G}_{011}-\frac{2}{9}\chi_{1+}\chi_{2-}d{\cal G}_{010},
\nonumber\\
dC_{\eta\eta} &=&\chi_{2+}^2d{\cal G}_{02}+\frac{1}{9}\chi_{1-}^2d{\cal G}_{011}-\frac{2}{9}\chi_{1-}\chi_{2+}d{\cal G}_{010}
\nonumber\\
&& +\chi_{P}^2d{\cal G}_{08}
-\chi_{P}\chi_{R}d{\cal G}_{07}+\chi_{R}^2d{\cal G}_{09}, \nonumber\\
\eee
with 
\bee\label{angcoef} 
d{\cal G}_{0k}&=&d\cos{\theta}d\tilde{\varepsilon}_1
\frac{G_{\beta}^4m_e^2 Q_{\beta\beta}}{64\pi^5 R^2}\nonumber\\
&\times& \left( h_{0k}(\varepsilon_1, \varepsilon_2, R) \cos{\theta}
+g_{0k}(\varepsilon_1, \varepsilon_2, R)\right)
p_1p_2\varepsilon_1\varepsilon_2 \nonumber \\
&\equiv& a_0^k +a_1^k \cos{\theta}
\qquad k=1,2,\cdots, 11 . \nonumber\\
\eee
Here, $a^k_0$ and $a^k_1$ are angular correlation coefficients. ${\cal G}_{0k} = \ln{2}~ G_{0k}$.

The differential decay rate can be written as  
\begin{equation}
\frac{d\Gamma}{d\cos{\theta}d\tilde{\varepsilon}_1}= a_0\lf 1 +a_1/a_0 \cos{\theta}\rh.
\end{equation}
Here $a_1/a_0$ is the energy-dependent angular correlation coefficient, which depends
also on the chosen limiting case for lepton number violating parameters. 

\begin{center}
  \begin{figure}[!t]
  \includegraphics[width=1.0\columnwidth]{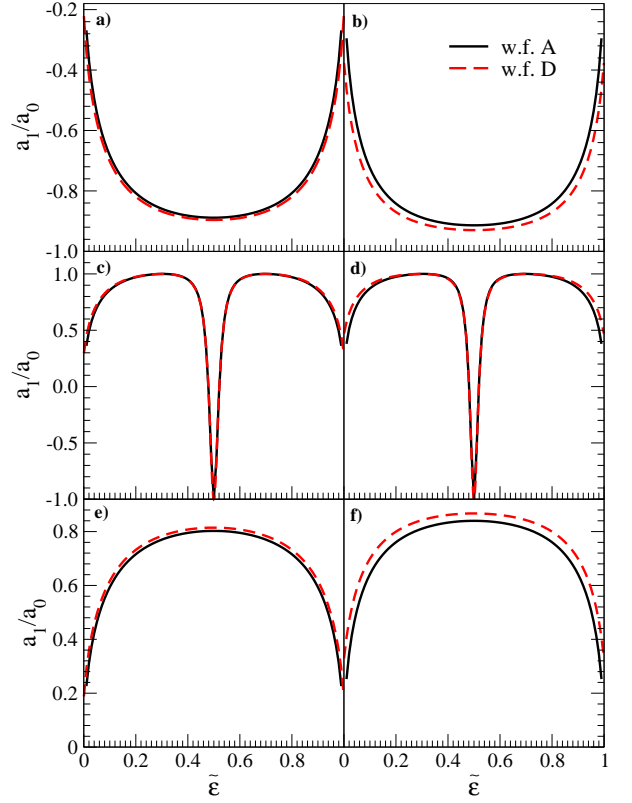}
  \caption{The angular correlation factor (see Eq. (\ref{angcoef})) 
    vs. the electron energy ${\tilde \varepsilon}$ (${\tilde \varepsilon}=(\varepsilon - m_e)/Q_{\beta\beta}$) 
    for $0\nu\beta\beta$-decay of $^{76}$Ge (left panels a), c) and e))
    and $^{136}$Xe (right panels b), d) and f)). Results are presented
    for $|m_{\beta\beta}|^2$ (panels a) and b)); $\langle \lambda \rangle = \langle \eta \rangle = 0$),
    $|\langle \lambda \rangle|^2$ (panels c) and d)); $m_{\beta\beta} = \langle \eta \rangle = 0$),
    $|\langle \eta \rangle|^2$ (panels e) and f)); $m_{\beta\beta} = \langle \lambda \rangle = 0$)
    terms. An approximate electron wave functions (w.f. A) and exact Dirac wave
    functions with finite nuclear size and electron screening (w.f. D) are considered.
    Nuclear matrix elements calculated within the ISM \cite{cau96} are used in calculations.} 
\label{fig:acf}
\end{figure}
\end{center}

In Fig. \ref{fig:endis} the single electron spectra normalized to the total decay rate are shown 
as function of the electron energy ${\tilde \varepsilon}$ for the $0\nu\beta\beta$-decay
of $^{76}$Ge and $^{136}$Xe due to non-vanishing $m_{\beta\beta}$, $\langle\lambda\rangle$ and
$\langle\eta\rangle$. This quantity, ideally accessible experimentally,
depends only very weakly on the chosen isotope, set of calculated nuclear
matrix elements, and whether standard or improved description of electron wave functions
is used. The different characteristics of these three limiting cases provide
a possibility to identify which of the parameters is responsible for
$0\nu\beta\beta$-decay.

\begin{center}
\begin{figure}[!t]
  \includegraphics[width=1.05\columnwidth]{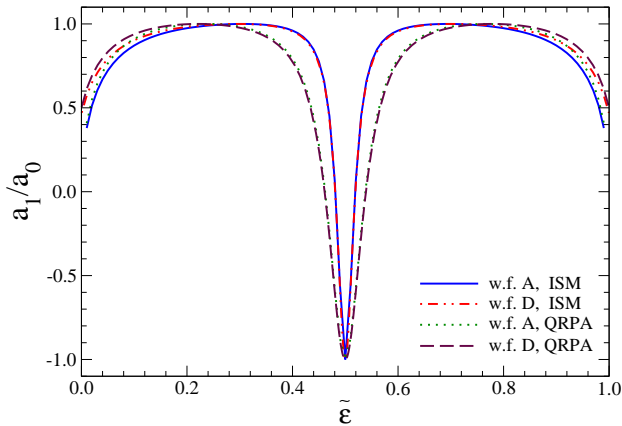} 
  \caption{The angular correlation factor (see Eq. (\ref{angcoef})) 
    vs. the electron energy ${\tilde \varepsilon}$ (${\tilde \varepsilon}=(\varepsilon - m_e)/Q_{\beta\beta}$) 
    for $0\nu\beta\beta$-decay of $^{136}$Xe. Results are presented
    for $|\langle \lambda \rangle|^2$ term ($m_{\beta\beta} = \langle \eta \rangle = 0$).
    An approximate electron wave functions (w.f. A) and exact Dirac wave
    functions with finite nuclear size and electron screening (w.f. D) are considered.
    Nuclear matrix elements calculated within the ISM \cite{cau96} and the QRPA \cite{muto89}
    are used in calculations.} 
\label{fig:osve}
\end{figure}
\end{center}

In Fig. \ref{fig:osve} the angular correlation factors $a_1/a_0$ are presented
as function of the electron energy ${\tilde \varepsilon}$ for the $0\nu\beta\beta$-decay
of $^{76}$Ge and $^{136}$Xe due to non-vanishing $m_{\beta\beta}$, $\langle\lambda\rangle$ and
$\langle\eta\rangle$. The ISM nuclear matrix elements are considered. The results
slightly depends on the type of electron wave functions and manifest similar behavior
for both isotopes. In Fig. \ref{fig:osve} the $a_1/a_0$ behavior in detail
for the $\langle\lambda\rangle$ limiting case is shown. Note that results 
in this case are affected more significantly by the choice of nuclear matrix elements
and that the choice of electron wave functions
changes $a_1/a_0$  only slightly for small and large values of electron energy.

%%%%%%%%%%%%%%%%%%%%%%%%%%%%%%%%%%%%%%%%%%%%%%%%%%%%%%%%%%%%%%%%%%%%%%%%%%%%%%%%%%%%%%%%%%%%%%%%%%%%
%
\section{Summary and Conclusions}
%
%%%%%%%%%%%%%%%%%%%%%%%%%%%%%%%%%%%%%%%%%%%%%%%%%%%%%%%%%%%%%%%%%%%%%%%%%%%%%%%%%%%%%%%%%%%%%%%%%%%%

It is often assumed that if and when the neutrinoless double beta decay ($0\nu\beta\beta$) is observed, it will be caused by the exchange of a virtual light
Majorana neutrino and its decay rate will be proportional to the square of the effective Majorana mass $m_{\beta\beta}$. 
It would be possible, therefore, to constrain or determine the magnitude of this fundamental parameter based on experimental limits or values
of the decay half-life. However, that is not the only
possibility. Many other manifestations of the `Physics beyond the Standard Model' that would cause the  $0\nu\beta\beta$ decay
were considered in the past. Among them the possibility of the right-handed lepton and/or hadron currents that could
perhaps compete with the mass mechanism was often discussed, see e.g. \cite{Doies,tomoda}. Such possibility arises naturally in the left-right
symmetric models. In that case the $0\nu\beta\beta$-decay half-life will depend not only on the $m_{\beta\beta}$ but also, perhaps dominantly,
on the parameters that characterize the right-handed currents, denoted $\langle \lambda \rangle$ and $\langle \eta \rangle$ here.

When it is assumed that the right-handed currents exist, the $0\nu\beta\beta$-decay half-life can be expressed as a sum of products
of the phase-space factors, nuclear matrix elements, and the fundamental parameters that characterize the new physics. In this
work the particular emphasis is on the reformulation of this relation, on a careful derivation of all terms in that expression, and on the new
and more general  evaluation of the phase-space factors. The phase-space factors depend on the wave functions of the emitted electrons,
and various approximations were used in the past in their calculation. We use here the exact solutions of the Dirac equation for the
$s_{1/2}$ and $p_{1/2}$ electron states, solving the Dirac equation in the potential that includes the nuclear finite size and the electron screening.
The possible approximations to this problem are analyzed and discussed, and in particular it is shown that using 
just the first order expansion in $r$ in order to include the
nuclear finite size with the sufficient accuracy, is not enough. Complete table of accurate phase-space factors for nuclei of interest is given.
Compared with the treatment that uses only the first terms in the $r$ expansion (denoted as approximation A in this work), the exact 
phase space factors (approximation D) are smaller, in particular in the heavier nuclei ($^{130}$Te and $^{150}$Nd) the reduction is
$\sim$ 30 \% or even more. 

It is also often assumed that the nucleons involved in the $0\nu\beta\beta$-decay are essentially at the nuclear surface, hence the phase space
factors are evaluated with the electrons placed at $r = R$. The adequacy of that assumption was not tested until now. Here it is shown, see
Fig. \ref{fig:rdep} and Table \ref{Tab.Priemer}, that it is a reasonable assumption, even though an increase in the phase-space factors by a few
percent in the heavier candidate nuclei is expected. 

 Having the full set of the phase-space factors, it is possible by combining them with the full set of nuclear matrix elements evaluated elsewhere,
 to obtain simultaneous or separate limits for the fundamental parameters $m_{\beta\beta}$ and those associated with the right-handed
 currents   $\langle \lambda \rangle$ and $ \langle \eta \rangle$. It turns out that again the difference between the previously used
 approximation A (just the first term in the expansion in $r$) and the more exact treatment (exact Dirac electron wave functions with the
 nuclear radius $R$ and electron screening) in the final limits is relatively benign in $^{76}$Ge, enlarging the limits on the fundamental
 parameters only by about 5\%. However, in the heavier nucleus $^{136}$Xe the effect is larger, 10-15\%.

 It is well known that by convincingly determining the $0\nu\beta\beta$ half-life one would obviously show that the total lepton number is
 not a conserved quantity. However, that determination by itself will be insufficient to decide which of the possible mechanism is responsible
 for the decay. If, in addition, the single electron energy spectra, and the angular distribution of the emitted electrons, could be detected,
 it will help substantially in that task.  If one of the possible parameters, $m_{\beta\beta}$, $\langle \lambda \rangle$, or $\langle \eta \rangle$
 dominates, the single particle spectra and angular correlations will be a decisive tool to determine the mechanism. Formulas that determine
 these quantities and the corresponding phase space factors are shown here. In that case the exact treatment
 of nuclear size makes only little difference.

%%%%%%%%%%%%%%%%%%%%%%%%%%%%%%%%%%%%%%%%%%%%%%%%%%%%%%%%%%%%%%%%%%%%%%%%%%%%%%%%%%%%%%%%%%%%%%%%%%%%
%%%%%%%%%%%%%%%%%%%%%%%%%%%%%%%%%%%%%%%%%%%%%%%%%%%%%%%%%%%%%%%%%%%%%%%%%%%%%%%%%%%%%%%%%%%%%%%%%%%%
\acknowledgments

This work is supported in part by the VEGA Grant Agency
of the Slovak Republic under Contract No. 1/0876/12, by Slovak Research and Development Agency under Contract
No. APVV-14-0524, and by the Ministry
of Education, Youth and Sports of the Czech Republic under Contract No. LM2011027.

%%%%%%%%%%%%%%%%%%%%%%%%%%%%%%%%%%%%%%%%%%%%%%%%%%%%%%%%%%%%%%%%%%%%%%%%%%%%%%%%%%%%%%%%%%%%%%%%%%%%
%

%
%%%%%%%%%%%%%%%%%%%%%%%%%%%%%%%%%%%%%%%%%%%%%%%%%%%%%%%%%%%%%%%%%%%%%%%%%%%%%%%%%%%%%%%%%%%%%%%%%%%%


\begin{thebibliography}{99}

\bibitem{schechter} J. Schechter and J.W.F. Valle, Phys. Rev. D {\bf 25}, 2591 (1982).

\bibitem{review} J.D. Vergados, H. Ejiri, and F. \v Simkovic, Rep. Prog. Phys. {\bf 71},
   106301 (2012).

\bibitem{iachello} J. Kotila and F. Iachello, Phys. Rev. C {\bf 85}, 034316 (2012).

\bibitem{Pati} J.C. Pati and A. Salam, Phys. Rev. D {\bf 10}, 275 (1974);
  R. Mohapatra and J.C. Pati, Phys. Rev. D {\bf 11}, 2558 (1975).
  
\bibitem{Mohap}
  G. Senjanovic and R.N. Mohapatra, Phys. Rev. D {\bf 12}, 1502 (1975);
  R. N. Mohapatra and G. Senjanovic, Phys. Rev. Lett. {\bf 44}, 912 (1980);
  Phys. Rev. D {\bf 23}, 165 (1981).
 
\bibitem{Doies} M. Doi, T. Kotani and E. Takasugi,  Prog. Theor. Phys. Supplement  {\bf 83}, 1 (1985).

\bibitem{tomoda} T. Tomoda,  Rep. Prog. Phys.  {\bf 54}, 53 (1991).

\bibitem{Tello} V. Tello, M. Nemev\v sek, F. Nesti, G. Senjanovi\'c and F. Vissani,
  Phys. Rev. Lett.  {\bf 106}, 151801 (2011).

\bibitem{Nemevsek} 
  M. Nemev\v sek, F. Nesti, G. Senjanovi\'c and V. Tello,
  arXiv:1112.3061 [hep-ph].

\bibitem{Barry13} J. Barry and W. Reodejohann,  JHEP {\bf 1309}, 153 (2013).

\bibitem{Dev} P.S. Bhupal Dev, S. Goswami and M. Mitra,
     Phys. Rev. D {\bf 91}, 113004 (2015). 
  
\bibitem{deppisch} F.Deppisch, J. Harz, M. Hirsch, Wei-Chih Huang, H. P\"as,
  arXiv:1503.04825[hep-ph].

  
\bibitem{rose} M.E. Rose, {\it Relativistic electron theory}, New York - London, 
  John Wiley and Sons, 1961.
 
\bibitem{landau} V.B. Beresteckij, E.M. Lifshitz, L.P. Pitaevskij,
  Volume IV, ``Quantum electrodynamics'', Nauka, Moskva 1989.  

\bibitem{radial} F. Salvat, J.M. Fernandez-Varea, W. Williamson Jr.,
     Comput. Phys. Commun. {\bf 90}, 151 (1995).

\bibitem{esposito} S. Esposito, Am. J. Phys. {\bf 70}, 852 (2002).

\bibitem{Xing} Zhi-zhong Xing, Phys. Rev. D {\bf 85}, 013008 (2012).


\bibitem{Stefanik} D. \v Stef\' anik, R. Dvornick\' y and F. \v Simkovic, Nuclear Theory {\bf 33}, 115 (2014).

\bibitem{lisi} 	G.L. Fogli, E. Lisi, A. Marrone, D. Montanino, A. Palazzo, Phys. Rev. D {\bf 86}, 013012 (2012). 

\bibitem{Simdva} F. \v Simkovic, et al., Phys. Rev. C {\bf 60}, 055502 (1999). 

\bibitem{vadimp} F. \v Simkovic, V. Rodin, A. Faessler, and P. Vogel, Phys. Rev. C {\bf 87}, 045501 (2013). 

\bibitem{muto89} K. Muto, E. Bender and H.V. Klapdor, Z. Phys. A {\bf 334}, 187 (1989).

\bibitem{cau96}
 E. Caurier, F. Nowacki, A. Poves and J. Retamosa, Phys. Rev. Lett. {\bf 77}, 1954 (1996).

\bibitem{cau08}
 E. Caurier, F. Nowacki, A. Poves and J. Retamosa, Eur. Phys. J.A {\bf 36}, 195 (2008).

\bibitem{gelimit}
The  GERDA Collaboration, M. Agostini et al., Phys. Rev. Lett. {\bf 111}, 122503 (2013).

\bibitem{xelimit}
The KamLAND-Zen Collaboration, A. Gando et al., Phys. Rev. Lett. {\bf 110}, 062502 (2013);
The EXO collaboration, M. Auger et al.,  Phys. Rev. Lett. {\bf 109}, 032505 (2012).

\bibitem{Super} The SuperNEMO Collaboration, R. Arnold et al., 
  Eur. Phys. J. C {\bf 70}, 927 (2010). 

\bibitem{Next} The NEXT Collaboration, V. Alvarez et al., arXiv:1106.3630 [physics.ins-det].
  

\end{thebibliography}
\end{document}